\documentclass[lettersize,journal]{IEEEtran}

\usepackage{amsmath,amsfonts}
\usepackage{xfrac}
\usepackage{array}
\usepackage{stfloats}
\usepackage{graphicx}
\usepackage{cite}
\usepackage[breaklinks]{hyperref}
\hypersetup{
    colorlinks=true,
    linkcolor=blue,
    citecolor=blue,
    urlcolor=cyan,
  }
\usepackage{graphicx}
\usepackage{xspace} 
\usepackage{subcaption}
\usepackage{multirow}
\usepackage{booktabs}
\usepackage{caption}
\usepackage{csquotes}
\usepackage{ifthen}
\def\BibTeX{{\rm B\kern-.05em{\sc i\kern-.025em b}\kern-.08em
    T\kern-.1667em\lower.7ex\hbox{E}\kern-.125emX}}
\usepackage{balance}
\usepackage{pifont}

\usepackage[binary-units]{siunitx}

\usepackage[capitalize]{cleveref}
\crefformat{section}{\S#2#1#3}
\crefformat{subsection}{\S#2#1#3}
\crefformat{subsubsection}{\S#2#1#3}
\crefname{figure}{Fig.}{Fig.}
\crefformat{figure}{Fig.\,#2#1#3}

\usepackage{bm}

\newcommand{\fakepar}{}
\newcommand{\boldpar}[1]{\vspace{0.3mm}\noindent\textbf{#1}}
\newcommand{\italicpar}[1]{\vspace{0.3mm}
  \noindent\textit{#1}}

\newcommand{\seq}[1]{\ensuremath{\,{#1}\,}}

\newcommand{\sysname}{CLoc}
\newcommand{\papername}{\textsc{\sysname}\xspace}
\newcommand{\cloc}{{\papername}\xspace}
\newcommand{\clocp}{\textsc{\sysname}\ensuremath{_\mathrm{P}}\xspace}
\newcommand{\clocs}{\textsc{\sysname}\ensuremath{_\mathrm{S}}\xspace}
\newcommand{\clocrs}{\textsc{\sysname}\ensuremath{_\mathrm{RS}}\xspace}

\newcommand{\challengename}{Problem\xspace}
\newcommand{\methodname}{Our technique\xspace}
\newcommand{\methodnames}{Our techniques\xspace}
\newcommand{\baselinename}{Baseline\xspace}

\newcommand{\challengepar}[1]{\boldpar{\challengename: {#1}.}\xspace}
\newcommand{\baselinepar}[1]{\boldpar{\baselinename: {#1}.}\xspace}
\newcommand{\methodpar}[1]{\boldpar{\methodname: {#1}.}\xspace}
\newcommand{\methodpars}[1]{\boldpar{\methodnames: {#1}.}\xspace}
\newcommand{\configpar}{\boldpar{Evaluation.}\xspace}
\newcommand{\evalpar}{\boldpar{Evaluation.}\xspace}

\usepackage{paralist}
\setdefaultenum{\em i)}{(a)}{1.}{A.}

\newcommand{\envtut}{University of Trento\xspace}

\newcommand{\envitihallway}{\textsc{tug}\xspace}
\newcommand{\envtuthallway}{\textsc{cloves}\xspace}
\newcommand{\tug}{\envitihallway}
\newcommand{\cloves}{\envtuthallway}

\newcommand{\chorus}{\textsc{Chorus}\xspace}
\newcommand{\snaploc}{\textsc{SnapLoc}\xspace}

\newcommand{\ie}{\textit{i.e.,}\xspace}
\newcommand{\eg}{\textit{e.g.,}\xspace}
\newcommand{\vs}{\textit{vs.\,}}
\newcommand{\wrt}{w.r.t.\xspace}
\newcommand{\perval}[1]{\ensuremath{#1}\%\xspace}

\DeclareSIUnit{\dB}{dB}
\DeclareSIUnit{\dBm}{dBm}
\DeclareSIUnit{\dBi}{dBi}

\newcommand{\sstrw}{\mbox{SS-TWR}\xspace}
\newcommand{\sstwr}{\sstrw}

\newcommand{\tdoa}{TDoA\xspace}
\newcommand{\ultdoa}{\mbox{UL-TDoA}\xspace}
\newcommand{\dltdoa}{\mbox{DL-TDoA}\xspace}

\newcommand{\ciri}{\ensuremath{\bm{C}_i}\xspace}
\newcommand{\cirinext}{\ensuremath{\bm{C}_{i+1}}\xspace}

\newcommand{\kmax}{\ensuremath{k_\mathrm{max}}\xspace}

\newcommand{\posest}{\ensuremath{\hat{\bm p}}\xspace}
\newcommand{\posexp}{\ensuremath{\bm p}\xspace}
\newcommand{\tdoaestv}{\ensuremath{\Delta\hat{\bm t}}\xspace}
\newcommand{\tdoaexpv}{\ensuremath{\Delta {\bm t}}\xspace}
\newcommand{\tdoaest}[1]{\ensuremath{\Delta\hat{t}_{#1}}\xspace}
\newcommand{\tdoaexp}[1]{\ensuremath{\Delta{t}_{#1}}\xspace}

\newcommand{\seg}{segment\xspace}
\newcommand{\segs}{segments\xspace}
\newcommand{\sect}{sub-segment\xspace}
\newcommand{\sects}{sub-segments\xspace}

\newcommand{\commaIfNotEmpty}[1]{\expandafter\ifx\expandafter\relax\detokenize{#1}\relax\else{,#1}\fi}

\newcommand{\pollpktstr}{poll}
\newcommand{\resppktstr}{resp}

\newcommand{\pollpkt}{\textsc{\pollpktstr}\xspace}
\newcommand{\resppkt}{\textsc{\resppktstr}\xspace}
\newcommand{\resp}{\resppkt}
\newcommand{\poll}{\pollpkt}
\newcommand{\dspktstr}{ds}
\newcommand{\dspkt}{\textsc{\dspktstr}\xspace}

\newcommand{\ttxpoll}{\ensuremath{t^{\mathrm{tx}}_{\mathrm{\pollpktstr}}}\xspace}
\newcommand{\trx}[1]{\ensuremath{t^{\mathrm{rx}}_{\mathrm{\resppktstr},{#1}}}\xspace}
\newcommand{\ttx}[1]{\ensuremath{t^{\mathrm{tx}}_{\mathrm{\resppktstr},{#1}}}\xspace}
\newcommand{\trxpoll}[1]{\ensuremath{t^{\mathrm{rx}}_{\mathrm{\pollpktstr},{#1}}}\xspace}

\newcommand{\trxds}[1]{\ensuremath{t^{\mathrm{rx}}_{\mathrm{\dspktstr},{#1}}}\xspace}

\newcommand{\schedres}{\ensuremath{\lambda}\xspace}
\newcommand{\drift}[1]{\ensuremath{\gamma_\mathit{#1}}\xspace}
\newcommand{\driftest}[1]{\ensuremath{\hat{\gamma}_{#1}}\xspace}

\newcommand{\fcbase}{\ensuremath{f}\xspace}
\newcommand{\fcnew}{\ensuremath{f^\prime}\xspace}
\newcommand{\fcoff}{\ensuremath{\Delta f}\xspace}

\newcommand{\ronew}{\ensuremath{\schedcorr^\prime}\xspace}

\newcommand{\fpsample}{\ensuremath{p_i}\xspace}
\newcommand{\fpsampleh}{\ensuremath{p_{i,h}}\xspace}
\newcommand{\fpsamplel}{\ensuremath{p_{i,l}}\xspace}

\newcommand{\tprepare}{\ensuremath{T_\mathrm{\resppktstr}}\xspace}
\newcommand{\tresp}{\tprepare}
\newcommand{\tdetune}{\ensuremath{T_\mathrm{det}}\xspace}
\newcommand{\tdetunep}{\ensuremath{\tdetune^\prime}\xspace}

\newcommand{\tshift}{\ensuremath{\delta}\xspace}
\newcommand{\tshifti}{\ensuremath{\delta_i}\xspace}
\newcommand{\tround}{\ensuremath{T_\mathrm{round}}\xspace}
\newcommand{\tds}{\ensuremath{T_\mathrm{ds}}\xspace}

\newcommand{\tkso}{\ensuremath{\tau}\xspace}

\newcommand{\wnoise}{\ensuremath{w_\mathrm{empty}}\xspace}

\newcommand{\truntime}{\ensuremath{T_\mathrm{run}}\xspace}
\newcommand{\tproc}{\ensuremath{T_\mathrm{proc}}\xspace}

\newcommand{\schederr}{\ensuremath{\varepsilon}\xspace}

\newcommand{\schedcorr}{\ensuremath{\rho}\xspace}

\newcommand{\cirp}{\ensuremath{\mathit{CIR}_\mathrm{P}}\xspace}
\newcommand{\cirs}{\ensuremath{\mathit{CIR}_\mathrm{S}}\xspace}

\newcommand{\CIRone}{\ensuremath{\mathit{CIR}_\mathit{S1}}\xspace}
\newcommand{\CIRtwo}{\ensuremath{\mathit{CIR}_\mathit{S2}}\xspace}

\newcommand{\doublesync}{DS\xspace}

\newcommand{\thold}{\ensuremath{T_\mathrm{hold}}\xspace}
\newcommand{\mlo}{\ensuremath{m_\mathrm{l}}\xspace}
\newcommand{\mhi}{\ensuremath{m_\mathrm{h}}\xspace}
\newcommand{\sback}{\ensuremath{s_\mathrm{b}}\xspace}
\newcommand{\trlo}{\ensuremath{\peakth_l}\xspace}
\newcommand{\trhi}{\ensuremath{\peakth_h}\xspace}

\newcommand{\trfp}{\ensuremath{\peakth^{'}_{i}}\xspace}
\newcommand{\trad}{\trfp}
\newcommand{\peakth}{\ensuremath{\theta}\xspace}
\newcommand{\peakthi}{\ensuremath{\theta_i}\xspace}
\newcommand{\sigman}{\ensuremath{\sigma_\mathrm{n}}\xspace}

\newcommand{\peaki}{\ensuremath{p_i}\xspace}

\newcommand{\interptext}{\ensuremath{\mathrm{int}}}
\newcommand{\peakiint}{\ensuremath{\peaki^\interptext}\xspace}

\newcommand{\peakipul}{\ensuremath{\peaki^\interptext}\xspace}
\newcommand{\peakidelta}{\ensuremath{\Delta p}\xspace}
\newcommand{\pcenter}{\ensuremath{k_{0}}\xspace}
\newcommand{\pfalling}{\ensuremath{\pcenter^{-}}\xspace}

\newcommand{\prising}{\ensuremath{\pcenter^{+}}\xspace}

\newcommand{\stsKey}{\ensuremath{\mathit{K\mspace{-4mu}ey}}\xspace}
\newcommand{\stsAKey}{\ensuremath{\stsKey_1}\xspace}
\newcommand{\stsBKey}{\ensuremath{\stsKey_2}\xspace}

\newcommand{\subcir}{\ensuremath{\gamma}\xspace}
\newcommand{\subciri}{\ensuremath{\gamma_i}\xspace}
\newcommand{\inonzero}{\ensuremath{{i\seq{\neq}0}}\xspace}
\newcommand{\subcirnonzero}{\ensuremath{\gamma_\inonzero}\xspace}
\newcommand{\Dmax}{\ensuremath{\Delta d_\mathrm{max}}\xspace}
\newcommand{\Smax}{\ensuremath{\Delta s_\mathrm{max}}\xspace}
\newcommand{\liorg}{\ensuremath{\mathit{LI}_\mathrm{org}}\xspace}
\newcommand{\liadj}{\ensuremath{\mathit{LI}_\mathrm{adj}}\xspace}
\newcommand{\qpf}{\ensuremath{\mathit{QPF}}\xspace}

\newcommand{\ltag}{\ensuremath{T}\xspace}
\newcommand{\lresp}{\ensuremath{R}\xspace}
\newcommand{\lrespi}{\ensuremath{R_i}\xspace}
\newcommand{\lrespref}{\ensuremath{R_\mathrm{ref}}\xspace}
\newcommand{\lanchor}{\lresp}
\newcommand{\lrefanchor}{\ensuremath{R_0}\xspace}
\newcommand{\lmainr}{\lrefanchor}

\newcommand{\pc}{\text{p50}\xspace}
\newcommand{\psf}{\text{p75}\xspace}
\newcommand{\pnt}{\text{p90}\xspace}
\newcommand{\pnn}{\text{p99}\xspace}
\newcommand{\dtxpw}{\ensuremath{\Delta\mathit{TX}_\mathrm{pwr}}\xspace}

\newcommand{\ieee}{IEEE\,802.15.4\xspace}
\newcommand{\ieeez}{IEEE\,802.15.4z\xspace}
\newcommand{\rmarker}{\texttt{RMARKER}\xspace}

\newcommand{\qorvofamily}{DW3000~family\xspace}
\newcommand{\dwfamily}{\qorvofamily}
\newcommand{\dwradio}{DW3110\xspace}
\newcommand{\dwmodule}{DWM3001C\xspace}
\newcommand{\dwplat}{DWM3001CDK\xspace}

\newcommand{\nonCtxDLTDoA}{staggered\xspace}
\newcommand{\nonctx}{\nonCtxDLTDoA}

\usepackage{tikz}
\newcommand*\circled[1]{\tikz[baseline=(char.base)]{
    \node[shape=circle,draw,inner sep=1pt] (char) {\bfseries\footnotesize #1};}}
\newcommand{\circledb}[1]{\circled{#1}}

\newboolean{authnotes}

\setboolean{authnotes}{true}
\setboolean{authnotes}{false}

\ifthenelse{\boolean{authnotes}}
{
\newcommand{\gp}[1]{\footnote{{\bf GP: #1}}}
\newcommand{\cab}[1]{\footnote{{\bf CA: #1}}}
\newcommand{\davide}[1]{\footnote{{\bf Davide: #1}}}
\newcommand{\matteo}[1]{\footnote{{\bf Matteo: #1}}}
\newcommand{\michael}[1]{\footnote{{\bf Michael: #1}}}
\newcommand{\enrico}[1]{\footnote{{\bf Enrico: #1}}}
\newcommand{\answer}[1]{{\color{blue}[#1]:}}

\newcommand{\notesmark}[1]{{\color{orange}#1\protect\footnotemark}}

\newcommand{\notestextcarlo}[1]{{\footnotetext{\color{orange}{\bf Notes Carlo: #1}}}}
\newcommand{\notestextgp}[1]{{\footnotetext{\color{orange}{\bf Notes GP: #1}}}}

\newcommand{\notescarlo}[2]{{\notesmark{#1}\notestextcarlo{#2}}}
\newcommand{\notesgp}[2]{{\notesmark{#1}\notestextgp{#2}}}

\newcommand{\memo}[1]{{\color{purple}
    \begin{compactitem}
    #1
    \end{compactitem}}}

 \usepackage{draftwatermark}
 \SetWatermarkLightness{0.85}
}{
\newcommand{\gp}[1]{}
\newcommand{\cab}[1]{}
\newcommand{\davide}[1]{}
\newcommand{\matteo}[1]{}
\newcommand{\michael}[1]{}
\newcommand{\enrico}[1]{}
\newcommand{\notescarlo}[2]{}
\newcommand{\notesgp}[2]{}

\newcommand{\memo}[1]{}

\newcommand{\answer}[1]{}
}

\crefrangeformat{figure}{Figs.~#3#1#4--#5#2#6}

\begin{document}
\title{Beyond Feasibility: Efficient and Robust On-Device Localization
  with UWB Concurrent Transmissions}

\author{
Enrico Soprana$\IEEEauthorrefmark{1}$, Michael Stocker$\IEEEauthorrefmark{1}$, Davide Vecchia, Matteo Trobinger, Carlo A. Boano, and Gian Pietro Picco \vspace{-7.00mm}
\thanks{$\IEEEauthorrefmark{1}$These authors contributed equally to this work. 
  Gian Pietro Picco, Enrico Soprana, Matteo Trobinger, and Davide
  Vecchia (\emph{firstname.lastname}@unitn.it) are with the
  Department of Information Engineering and Computer Science at the University of Trento, Italy.  Carlo Alberto Boano (\emph{cboano}@tugraz.at) and Michael
  Stocker (\emph{mstocker}@student.tugraz.at) are with the
  Institute of Technical Informatics at Graz University of Technology,
  Austria.}
}


\maketitle

\begin{abstract}
Concurrent transmissions (CTX) enable an ultra-wideband (UWB) tag to
range to, or self-localize from, multiple nodes with a single frame
reception, significantly improving update rate and energy efficiency
compared to traditional approaches.  Prior work demonstrated the
concept feasibility via prototypes whose real-world applicability is
however hindered by many limitations, notably including their
inability to directly execute on UWB tags and to cope with
interfering~UWB signals.

We~address these limitations in two ways. First, we redesign the core
CTX techniques to enable a fast and efficient execution on
resource-constrained MCUs without sacrificing accuracy. Second, we
leverage the new features in the \ieeez standard to achieve robustness
to interference from other UWB devices and enable tags to dynamically
select which anchors to use, two crucial and unexplored capabilities
in CTX-based systems.  We quantitatively evaluate our approach
experimentally in two real-world testbeds with different
characteristics.  We first analyze our new techniques individually,
offering an in-depth discussion of design choices and performance
trade-offs. Then,~we~embed~all our techniques into a CTX-based system
enabling a tag to~self-localize entirely on-device, unlocking gains up
to 44\% in runtime~and 85\% in energy efficiency compared to
traditional
approaches.


\end{abstract} 


\vspace{-1.00mm}
\section{Introduction}
\label{sec:intro}

\noindent
Ultra-wideband (UWB) has emerged as the cornerstone technology for a variety of location-aware IoT 
applications~\cite{UWBAlliance_Applications}.
\mbox{Exploiting} a unique physical layer, UWB radios estimate the channel impulse response (CIR) of an incoming frame 
to determine its time of arrival (ToA), \ie \mbox{RX timestamp}, with sub-\SI{}{\nano\s} resolution. 
This enables precise distance estimation (ranging) and localization, 
commonly using Time-of-Flight (ToF) or Time-Difference-of-Arrival (TDoA) measurements.
Various schemes exist to acquire these measurements, each offering a different balance 
across accuracy, reliability, channel occupancy, energy efficiency, scalability, and update rate.
Recent works have pushed the boundaries on several of these aspects at once by using \emph{concurrent transmissions} (CTX).

\boldpar{CTX for ranging and localization.}  Originally developed for
fast, reliable, and energy-efficient communication in narrowband
networks~\cite{zimmerling2020synchronous}, CTX have also been used in
UWB for the same purpose~\cite{weaver, flick}, but their impact goes
far beyond communication.  As first shown with concurrent
\mbox{ranging~\cite{corbalan2018concurrent,
    grosswindhager2018concurrent, heydariaan19R3, corbalan2020ultra}}, the RX
timestamps from different simultaneous transmitters can be derived
from the \emph{same} CIR, yielding multiple ToF measurements \emph{at
  once}.
This insight has later been applied to TDoA-based
localization 
in \chorus~\cite{corbalan2019chorus} and
\snaploc~\cite{grosswindhager2019snaploc}, in which the fixed nodes
(\emph{anchors}) of a UWB infrastructure transmit at the same time,
and any number of listening tags can self-localize with a
\emph{single} reception. 
Subsequent studies have demonstrated that CTX can also be effectively employed for angle-of-arrival (AoA) estimation~\cite{heydariaan20anguloc, smaoui21single}, as well as hybrid AoA–TDoA localization~\cite{tianyu20SingleAnchor,smaoui21anchor}.
In contrast, traditional ranging and
localization systems require \emph{distinct} receptions for
\emph{each} timestamp, giving CTX-based approaches clear advantages~(\cref{sec:background}--\ref{sec:ctxbasics}) in
terms of channel occupancy, energy efficiency, scalability, and update
rate---at least in
principle.  

\boldpar{Limitations of CTX-based approaches.}  
These works focused on demonstrating \emph{feasibility} rather than ensuring
robustness and real-world applicability. As a result, CTX-based
ranging and localization approaches \textit{are not yet ready for
  practical use}, mainly due to the following unaddressed challenges (\cref{sec:overview}).

\italicpar{Inability to operate on-device.}
Existing works execute on-device only the packet exchange: the crucial
steps of processing the CIR and extracting RX~timestamps occur offline or rely on external 
infrastructure. 
As a consequence, these algorithms cannot run on the battery-powered,
low-cost, resource-constrained tags common in UWB applications, \eg
robot coordination, contact tracing, or asset
tracking~\cite{yu2021robot, 6Fit}.
Enabling on-device execution demands new techniques that adapt, when
not completely rethink, the existing ones to yield accurate yet fast, lightweigth, and energy-efficient execution.

\italicpar{Susceptibility to real-world dynamics.}
Revisiting CTX-based operations is also necessary to push the state of the art beyond existing techniques, which are confined to specific setups and neglect robustness to real-world dynamics.  
For example, deriving RX timestamps from a CIR requires selecting a threshold to distinguish valid peaks from~noise~\cite{corbalan2020ultra, grosswindhager2019snaploc}. 
However, the choice of this threshold involves unexplored tradeoffs between sensitivity to weak signals and robustness against multipath. 
Similarly, the robust identification of concurrent
responders relies on artificial constraints  
such as reserving empty portions of the CIR~\cite{corbalan2020ultra}, which limits the number of supported anchors, the spatial coverage of the system, and resilience to multipath. 

\italicpar{Inability to cope with unwanted UWB signals.}~CTX-based 
ranging and localization schemes have so far been demonstrated
only in \emph{isolated} setups, \ie where no other UWB devices are
active. This approach yields clean and easily interpretable CIRs,
crucial to ascertain feasibility, but fails to reflect the complexity
of real-world scenarios, whose target areas may contain more anchors
than those involved in a CTX exchange, as well as UWB devices
belonging to other systems. In both cases, these ``external'' UWB
signals may interfere with CTX operation, which must therefore be
augmented with the ability to discriminate and select the relevant UWB
signals from unwanted ones. However, whether and how this important
capability can be achieved is currently unexplored.

\boldpar{Contributions.}  This work removes the limitations above, paving the way for the practical adoption of CTX-based UWB ranging and localization systems in real-world settings.

To address on-device operation, we present techniques~that enable full
CTX execution 
\textit{directly on off-the-shelf UWB devices}. We \mbox{revisit} in
depth all CTX constituents (\cref{sec:cloc-design}), from the exchange
of concurrent UWB packets to the CIR processing required to derive the
RX
timestamps. 
The latter step is especially challenging, as it must be fast and
compact enough to seize the benefits of CTX, yet it involves a
time-consuming extraction over SPI of the CIR and its
computation-heavy analysis on resource-constrained~MCUs.
Our techniques markedly reduce this overhead without sacrificing
accuracy and robustness. Moreover, we limit the required CIR data and
accelerate on-device execution via direct memory access (DMA) to
parallelize CIR extraction and analysis~(\cref{sec:ondevice}).

These novel techniques we propose are general and applicable to most
IEEE\,802.15.4~UWB radios, including the popular Qorvo~DW1000 adopted by
all existing CTX works. In contrast, we focus on the newer
Qorvo~DW3000, 
compliant to the \ieeez 
standard. 
This choice, \emph{unique in the CTX literature}, 
aligns our findings to the state~of~the~art in UWB and also provides the base to augment CTX operation with the ability to \textit{isolate and select} only the relevant UWB signals~(\cref{sec:overview}), via the \textit{Scrambled Timestamp Sequence} (STS). 
The latter, introduced in \ieeez, is a ciphered sequence derived from a known key, aimed at protecting against malicious manipulation of RX timestamps. 
By correlating with the expected STS, a UWB receiver can obtain a second, ``clean'' CIR that isolates the contributions of transmitters possessing the correct key while suppressing signals from unauthorized or unknown devices. 
We exploit this capability to provide \textit{responder selection} in CTX via a novel key switching scheme, which enables UWB devices to acquire RX timestamps only from concurrent transmitters \emph{of their choice}~(\cref{sec:concurrentsts}). 

We evaluate our techniques via \textit{extensive experiments in two real-world 
  testbeds}~(\cref{sec:methodology}), showing a consistent performance
despite their largely diverse geometry and environmental
characteristics. We initially focus on each individual technique,
analyzing in detail its design and performance tradeoffs and, when
available, comparing it against the state of the art.
Then, although our techniques are in principle applicable to any
CTX-based system for ranging and localization, we concretely
demonstrate their use in synergy via \cloc, a new \tdoa-based
localization system incorporating all of our findings, and the first
CTX-based system operating entirely
on-device~(\cref{sec:putting_things_together}).

Our results show that non-concurrent systems are up to 44\% slower and
85\% more energy-hungry than \cloc, depending on their configuration.
They also reveal that our techniques are so effective that the main
bottleneck is now the SPI interface, currently the only way to extract
the CIR from off-the-shelf devices. This suggests that faster
mechanisms in next-generation UWB radios will unlock further
improvements from our techniques.  After discussing these results and
their implications, we close with brief concluding remarks
(\cref{sec:ending}).


\section{Background and Related Work}
\label{sec:background}
\vspace{-0.50mm}

\begin{figure*}[t]
\centering
\vspace{-2.50mm}
\begin{subfigure}[b]{0.31\textwidth}
    \centering
    \includegraphics[width=\textwidth]{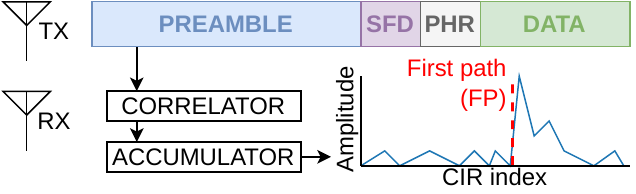}
    \vspace{-5.75mm}
    \caption{Base frame}
    \label{fig:basics-frames:single}
\end{subfigure}
\hfill
\begin{subfigure}[b]{0.29\textwidth}
    \centering
    \includegraphics[width=\textwidth]{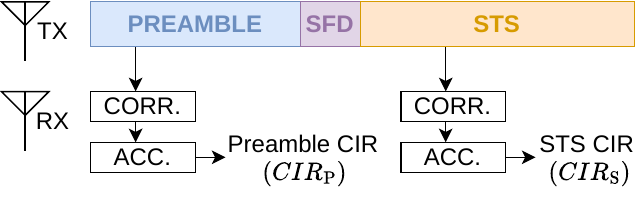}
    \vspace{-5.75mm}
    \caption{Frame with STS (\texttt{SP3} mode)}
    \label{fig:basics-frames:sts}
\end{subfigure} 
\hfill
\begin{subfigure}[b]{0.31\textwidth}
    \centering
    \includegraphics[width=\textwidth]{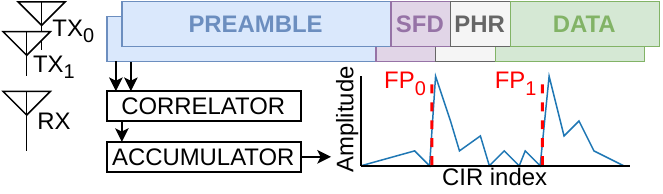}
    \vspace{-5.75mm}
    \caption{Concurrent frames}
    \label{fig:basics-frames:ctx}
\end{subfigure}
\vspace{-1.25mm}
\caption{Transmission and reception of \ieee UWB frames and CIR estimation process.}
\label{fig:basics-frames}
\vspace{-3.50mm}
\end{figure*}

\noindent 
We outline next key aspects of UWB
technology~(\cref{subsec:background_basics}) and review common
ranging and localization techniques, highlighting the benefits of
CTX-based
approaches~(\cref{subsec:background_schemes}). 

\vspace{-2.00mm}
\subsection{UWB basics and \ieeez enhancements}
\label{subsec:background_basics}
\vspace{-0.50mm}

\noindent
UWB is a low-power wireless technology using a large bandwidth,
either $>$\,$\SI{500}{\mega\hertz}$ or $>$\,$20\%$ of the center
frequency.  
Interest in UWB revived in 2013 with the release of the DW1000
radio~\cite{dw1000-manual}, offering compact size, high accuracy, and
low power consumption. 
Since then, many companies 
entered the UWB
market. 
Moreover, in 2019 Qorvo released the DW3000
radio~\cite{qorvo2021usermanual}, compliant with 
\ieeez. 

\fakepar All of the above are impulse UWB radios, encoding information
via very short pulses ($<$\,$\SI{2}{\nano\s}$).  This property
allows~a receiver to discern the signal direct \emph{first path (FP)}
from multipath components (MPCs), enabling UWB's key feature: the
ability to estimate with \emph{sub-ns} accuracy the ToA of a signal,
yielding \emph{decimeter-level} accuracy for distance estimation.  To
understand how, we must delve deeper into the UWB physical
layer~(PHY), standardized in
\ieee 
since 2007. 

\boldpar{Frame structure.}  A UWB frame
(\cref{fig:basics-frames:single}) consists of a preamble of variable
duration, followed by the start-of-frame delimiter (SFD) marking the
beginning of the modulated physical header (PHR) and the data payload.
The departure and arrival times of the first PHR symbol (\rmarker)
correspond to the TX and RX timestamps recorded by the radio; these
can be used to calculate the ToA and ToF. 

The UWB preamble consists of repeated symbols generated from a predefined pulse sequence with ideal periodic autocorrelation~properties, enabling reliable signal detection and synchronization. 
In this context, the Preamble \mbox{Acquisition} Chunk (PAC) defines how many preamble symbols are processed during the correlation-based acquisition process. It thus determines the granularity of the preamble detection procedure and can influence the robustness 
of frame synchronization. 

\boldpar{CIR estimate and RX timestamps.} 
After successful preamble acquisition, coherent receivers 
can estimate the CIR by correlating the received signal with the expected pulse pattern over the duration of the preamble. 
In the DW1000 and DW3000 radios, the CIR is accessible via an accumulator of \SI{4}{\kilo\byte} and \SI{6}{\kilo\byte}, respectively, containing approximately 1016 samples with a time resolution of
$\approx$\,$\SI{1}{\nano\s}$ each. 
The radio determines the FP of a signal as the first peak in the CIR
(\eg the dashed red line in \cref{fig:basics-frames:single}). 
In both DW1000 and DW3000 radios,
this initial, coarse RX timestamp is refined using a proprietary algorithm to obtain the final timestamp, whose resolution is $\approx$\,$\SI{15.65}{\pico\s}$, corresponding to a spatial resolution of
$\SI{4.69}{\mm}$.

\boldpar{Scrambled Timestamp Sequence (STS).} 
\emph{Any} UWB device transmitting on the same RF
channel near the receiver may introduce additional peaks in the CIR. If these are mistakenly considered during FP detection, they can cause significant errors in the RX timestamp (and thus distance) estimation. 

To mitigate this issue, the \ieeez~standard 
introduced an optional 
frame portion through which a receiver can access a CIR estimate
derived \emph{only} from the STS
(\cirs~in~\cref{fig:basics-frames:sts}) in addition to the base one
obtained from~the preamble (\cirp).  STS symbols, unlike preamble
ones, are not repeated and are based on cryptographic material shared
between transmitter and receiver. This allows a receiver to estimate
\cirs by correlating the received signal with~a secret and unpredictable template, improving resilience to intentional and
unintentional UWB interference during RX timestamping.

\ieeez defines four \emph{modes} with the STS: 
\begin{inparaenum}
\item omitted (\texttt{SP0},~\cref{fig:basics-frames:single}),
  yielding only \cirp;
\item inserted after the SFD (\texttt{SP1}); 
\item appended after the payload (\texttt{SP2}); and
\item sent \textit{instead} of the PHY header and payload
  (\texttt{SP3},~\cref{fig:basics-frames:sts}), minimizing frame
  length and collisions.
\end{inparaenum}
Since CTX approaches do not rely on data payloads, we always use
mode~\texttt{SP3}.

\begin{figure*}[b]
\centering 
\newlength{\figlen}
\setlength{\figlen}{1\textwidth}
\newlength{\figheight}
\setlength{\figheight}{1.3cm}

\vspace{-3mm}
\captionsetup[subfigure]{font=footnotesize,justification=centering}
\begin{subfigure}[t]{0.140897\figlen}
	\centering
	\includegraphics[height=\figheight, keepaspectratio]{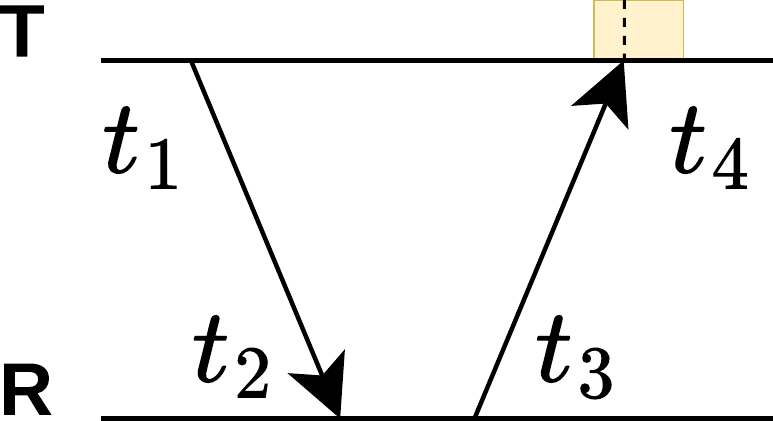}
	\caption{\sstwr}
	\label{fig:ss_twr}
\end{subfigure}%
\begin{subfigure}[t]{0.196885\figlen}
	\centering
	\includegraphics[height=\figheight, keepaspectratio]{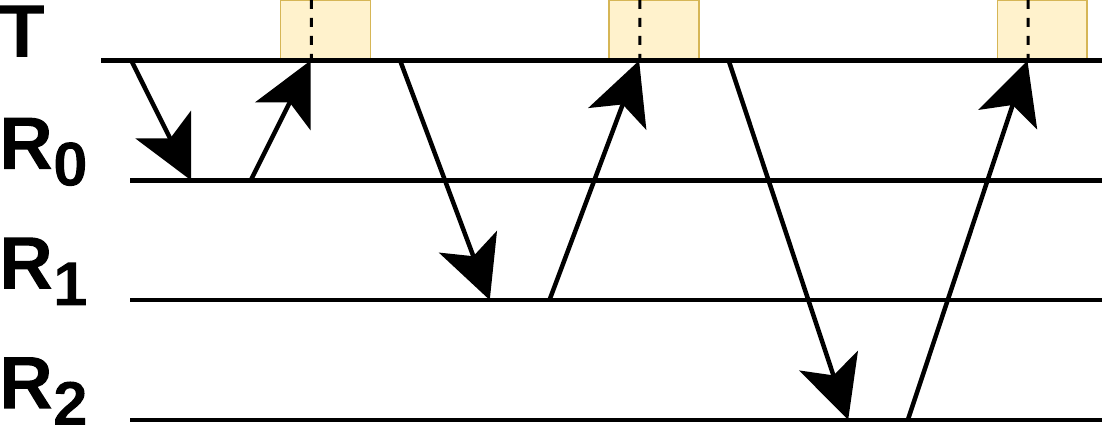}
	\caption{\sstwr\\(sequential)}
	\label{fig:ss_twr_seq}
\end{subfigure}%
\begin{subfigure}[t]{0.151650\figlen}
	\centering
	\includegraphics[height=\figheight, keepaspectratio]{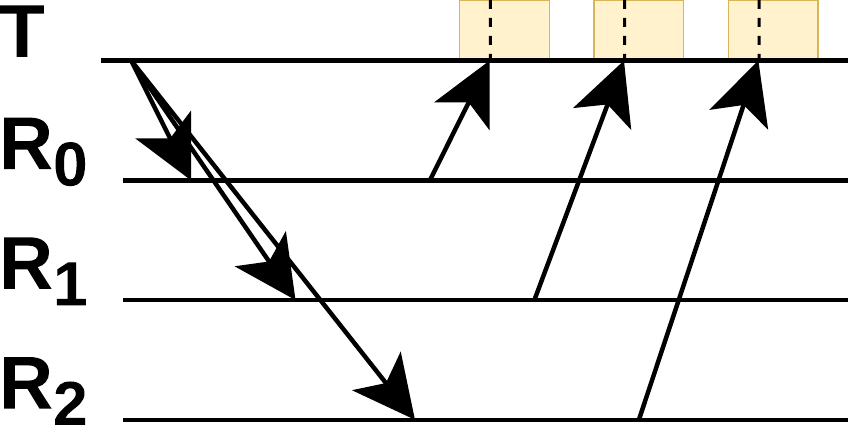}
	\caption{\sstwr\\(staggered)}
	\label{fig:ss_twr_staggered}
\end{subfigure}%
\begin{subfigure}[t]{0.116796\figlen}
	\centering
	\includegraphics[height=\figheight, keepaspectratio]{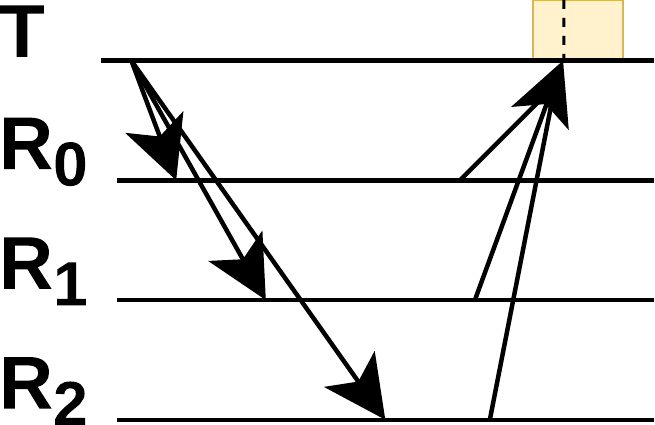}
	\caption{\sstwr\\(CTX)}
	\label{fig:ss_twr_ctx}
\end{subfigure}%
\begin{subfigure}[t]{0.151650\figlen}
	\centering
	\includegraphics[height=\figheight, keepaspectratio]{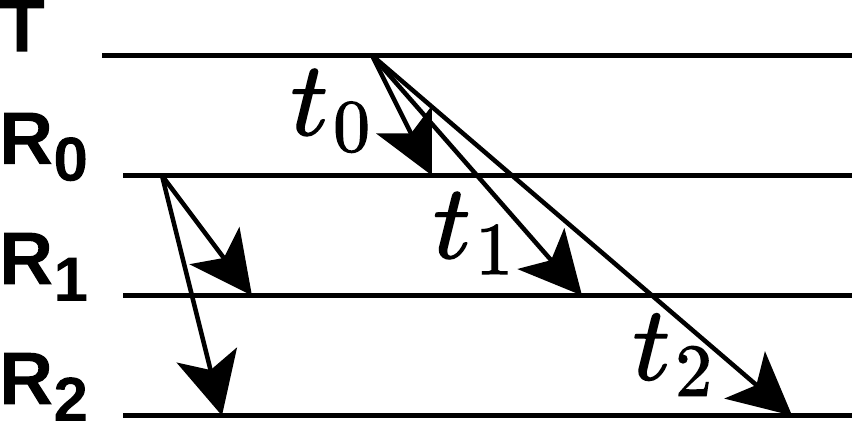}
	\caption{UL-TDoA}
	\label{fig:ul_tdoa}
\end{subfigure}%
\begin{subfigure}[t]{0.151650\figlen}
	\centering
	\includegraphics[height=\figheight, keepaspectratio]{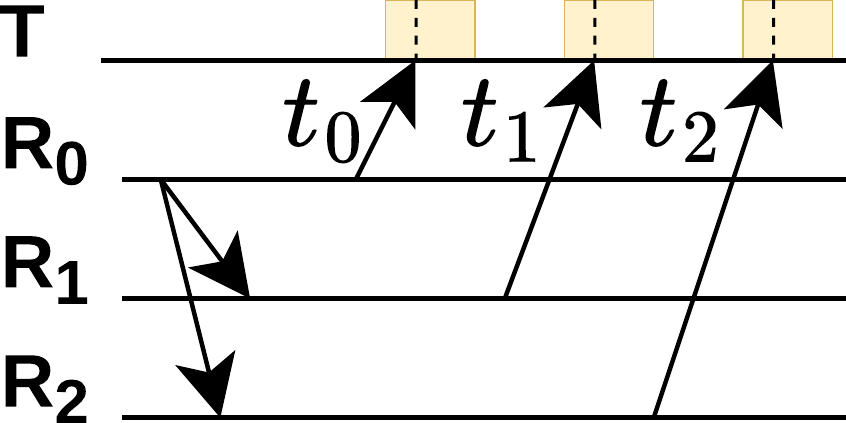}
	\caption{DL-TDoA}
	\label{fig:dl_tdoa}
\end{subfigure}%
\begin{subfigure}[t]{0.090471\figlen}
	\centering
	\includegraphics[height=\figheight, keepaspectratio]{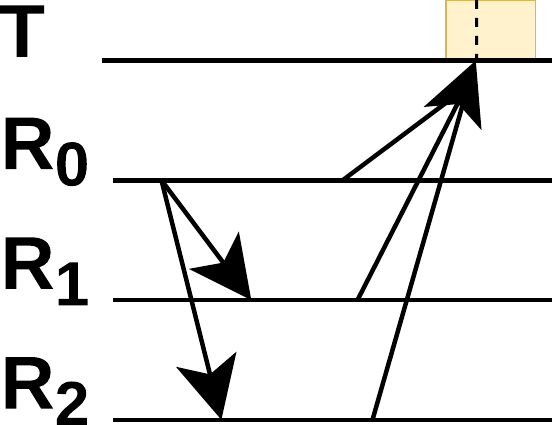}
	\caption{DL-TDoA\\(CTX)}
	\label{fig:dl_tdoa_ctx}
\end{subfigure}%

\vspace{-1.75mm}
\caption{Basic messaging schemes for \sstwr and \tdoa
  localization. The dashed lines within the received frames at the
  tag~\ltag denote RX events. In~(e)--(g), each responder $\lanchor_i$
  is assumed to be an anchor of a time-synchronized infrastructure.}
\label{fig:basics-schemes}
\vspace{-1.25mm}
\end{figure*}

\vspace{-0.50mm}
\subsection{UWB ranging and localization schemes}
\label{subsec:background_schemes}

\noindent
The accurate timestamping of UWB is exploited by ranging and
localization schemes; \cref{fig:basics-schemes} shows the most common
ones.

\boldpar{Ranging with a single responder.} In \emph{single-sided
  two-way ranging} (\sstwr, \cref{fig:ss_twr}), a tag \ltag sends a
\pollpkt packet to a responder \lresp, which replies with a \resppkt
packet. This round-trip exchange enables \ltag to estimate its
distance from \lresp as
\mbox{$d = c \times \mathit{ToF} = c \times
  \frac{1}{2}((t_4-t_1)-(t_3-t_2))$}, where $t_1$ and $t_3$ are TX
timestamps, $t_2$ and $t_4$ are RX timestamps (\ie ToA measurements),
and $c$ is the speed of light in air. 
Thanks to the ability of UWB
radios to accurately schedule transmissions, \lresp can embed the known transmission time $t_3$ into the \resppkt packet, providing \ltag 
with all necessary data to compute $d$.

\boldpar{Ranging with multiple responders.}  Several approaches exist for
enabling \ltag to target multiple responders, \eg nearby~tags  
or known anchors.  In the simplest one, \ltag polls each of~the
$N$ responders \emph{sequentially} (\cref{fig:ss_twr_seq}), thus requiring $2N$
messages over $N$ round-trip exchanges.
A~more efficient alternative is to broadcast a single \pollpkt packet and \emph{stagger} the responses (\cref{fig:ss_twr_staggered}); this reduces the amount of messages to $N\seq{+}1$,
yielding a shorter duration of the overall
exchange~\cite{patent_staggered,ntwr}. 

\boldpar{\tdoa-based localization.} The distance estimates acquired
with ranging enable the tag \ltag to \textit{self-localize} via
true-range multilateration, based on the known positions of enough 
responders (anchors)~\cite{buehrer18Handbook}.
Another option is to use \tdoa systems, which require the anchors to 
\emph{share} an accurate time reference. 
This can 
be achieved either out-of-band through wired clock distribution or in-band via
UWB~\cite{vecchia2019talla, tiemann2016atlas}. Localization then relies on the difference in
propagation time induced by the different anchor positions \wrt the~tag.
Two~classes~of~\tdoa schemes exist, depending on the role of
devices. In \textit{uplink}~\tdoa~(\ultdoa,~\cref{fig:ul_tdoa}), the
tag broadcasts a packet to all anchors, which record the RX timestamp. 
\ultdoa is inherently more scalable than TWR,
since a single packet transmission from the tag enables position
computation. Nevertheless, tag transmissions must be properly
scheduled to avoid collisions; moreover, many applications require tags to 
self-localize, 
\eg to preserve privacy. 
\textit{\mbox{Downlink}}~\tdoa~(\dltdoa, \cref{fig:dl_tdoa})
accounts for these concerns with a reversed scheme where anchors
transmit continuously according to a predefined order and the tag
locally computes its position based on the differences between RX
timestamps~\cite{puatru2023flextdoa}. \dltdoa hence preserves privacy while scaling to countless tags, since tags only \emph{receive} packets.

\boldpar{CTX-based approaches.} The use of CTX redefines the above
tradeoffs among scalability, update rate, energy consumption, and
self-localization, enabling new schemes where these conflicting
concerns can be reconciled effectively and efficiently. 
When multiple devices transmit concurrently, the radio correlation procedure at the receiver yields a CIR ideally containing one peak \emph{for each} of the overlapping signals~(two in~\cref{fig:basics-frames:ctx}). 
This property can be exploited in ranging and localization schemes that \emph{deliberately enforce} these simultaneous transmissions and, by processing the resulting CIR, extract from a \emph{single} reception event \emph{all} the RX timestamps, each associated to a signal peak. 
The reduction of radio activity yields direct benefits to update rate, energy consumption, and scalability.

This idea appeared first in \mbox{\emph{concurrent ranging}~\cite{corbalan2018concurrent, grosswindhager2018concurrent}}, a \mbox{\sstwr} scheme where \ltag broadcasts its
\pollpkt packet as in the staggered variant, but the responders~\lrespi~send~their~\resppkt~packet \emph{at the same time}~(\cref{fig:ss_twr_ctx}). 
Along the same lines,~CTX-based versions of \dltdoa localization were \mbox{devised~\cite{corbalan2019chorus,
grosswindhager2019snaploc}}, where anchors transmit simultaneously,
rather than one at a time,
upon receiving a synchronization
packet broadcast by an anchor \lrespref ($\lresp_0$
in~\cref{fig:dl_tdoa_ctx}) serving as time reference.

Despite their different goals, concurrent ranging and concurrent
\dltdoa schemes have the same structure, as shown by
Figs.\,\ref{fig:ss_twr_ctx} and \ref{fig:dl_tdoa_ctx}. The only
difference lies in the device initiating the CTX exchange: the
tag~\ltag~in~concurrent ranging \vs the reference anchor \lrespref in
concurrent \dltdoa. This observation enables next a \emph{unified}
presentation of the core concepts behind these CTX approaches~and~of
the 
challenges we tackle. Moreover, 
it also implies that \emph{our findings are applicable to both
  concurrent ranging and concurrent localization schemes}.


\section{Concurrent Transmissions: A Closer Look}
\label{sec:ctxbasics}

\begin{figure*}[!t]
    \vspace{-5mm}
    \centering
    \includegraphics[width=0.985\linewidth]{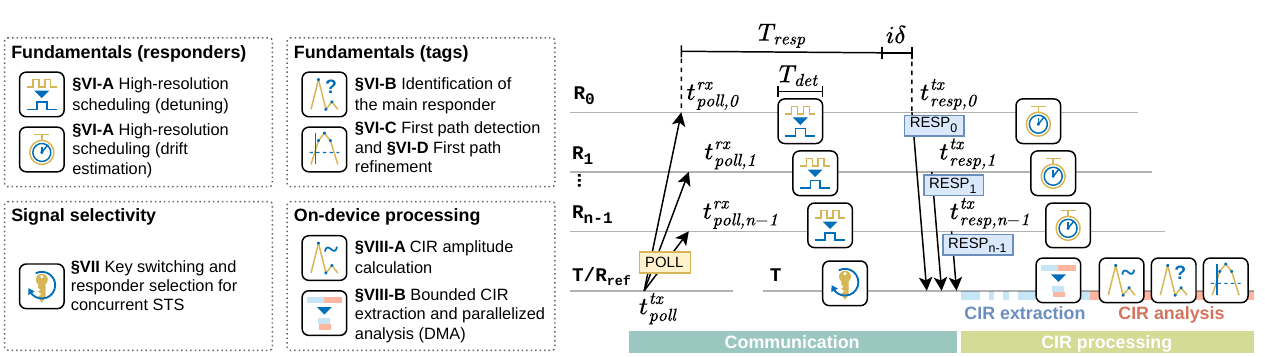}
    \vspace{-1.00mm}
    \caption{Overview of the techniques we propose to achieve robust on-device ranging and localization with UWB CTX. The concurrent exchange scheme shown on the right applies to both ranging and localization. It is used as a basis to evaluate each individual technique (\cref{sec:cloc-design}--\cref{sec:ondevice}), and to create \cloc, a CTX-based \dltdoa system embedding all our findings (\cref{sec:putting_things_together}).}
    \label{fig:scheme}
    \vspace{-1.75mm}
\end{figure*}

\noindent
\cref{fig:scheme} (right) illustrates the key concepts at the core of
both CTX-based ranging and localization approaches
(\cref{subsec:background_schemes}).

\begin{figure}[!b]
    \centering
    \vspace{-2mm}
    \setlength{\belowcaptionskip}{0pt}
    \setlength{\abovecaptionskip}{0pt}
    \setlength{\floatsep}{0pt}
    \setlength{\intextsep}{0pt}
    \setlength{\dbltextfloatsep}{0pt}
    \setlength{\dblfloatsep}{0pt}
    \includegraphics[width=\columnwidth]{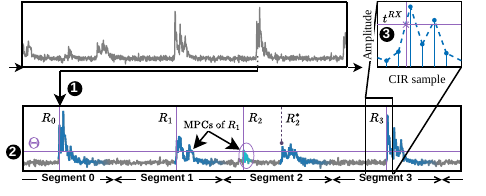}
    \includegraphics[width=0.99\columnwidth]{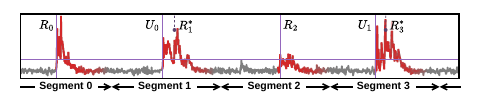}
    \begin{subfigure}[t]{\columnwidth}
        \centering
        \vspace{-46.9mm}
        \caption{\mbox{Input: Raw preamble CIR.}}
        \label{fig:stscir_raw}
    \end{subfigure}
    \begin{subfigure}[t]{\columnwidth}
        \centering
        \vspace{-30.5mm}
        \caption{\mbox{Main steps of CIR analysis.}}
        \label{fig:stscir_analysis}
    \end{subfigure}
    \begin{subfigure}[t]{\columnwidth}
        \centering
        \vspace{-15.75mm}
        \caption{\mbox{Impact of unwanted signals $U_i$.}}
        \label{fig:stscir_interfered}
    \end{subfigure}
    \vspace{-13mm}
    \caption{Analysis of a CIR obtained in the presence of four concurrent responders. 
    $R^*_i$ is the true FP associated to responder $i$; $R_i$ is the FP actually identified by the analysis.}
    \label{fig:stscir}
    \vspace{-1.25mm}
\end{figure}

The first phase is the \emph{communication} necessary to trigger the
transmission of the overlapping signals.
Depending on the scheme, an \emph{initiator}---either \ltag
  or \lrespref---begins the exchange by broadcasting a \pollpkt
packet at time \ttxpoll.  After a known delay, each responder \lrespi
transmits its \resppkt packet ($\resppkt_i$). At the tag~\ltag, the
reception of these concurrent packets generates a \emph{single} CIR
containing the overlapped signals from \lrespi
(\cref{fig:stscir_raw}). The next phase, the \emph{CIR processing},
encompasses the \emph{extraction} of the CIR from the radio via SPI
and its \emph{analysis} to yield the RX timestamps associated with
each responder. In this section, we focus exclusively on the preamble
CIR (\cirp), as it is the only one used in the CTX literature.

We discuss next the distinctive aspects concerning the use of CTX,
along with the techniques employed by existing works.

\boldpar{Scheduling \resppkt packets.} In non-concurrent approaches,
each \resppkt embeds its TX timestamp and sender~ID,
enabling direct ToF computation and sender association.  In CTX-based
approaches, this is not possible: the CIR at \ltag contains the
signals of \emph{all} responders, but the radio can decode \emph{only
  one data payload}. Moreover, simultaneous
transmissions may produce overlapping CIR peaks that become
indistinguishable.

Existing works~\cite{corbalan2020ultra,corbalan2019chorus,grosswindhager2019snaploc}
tackle these problems by temporally separating \resppkt transmissions via a TX delay
\begin{equation}
  \ttx{i} = \trxpoll{i} + \tprepare + i\tshift
\label{eq:ttx_rng}  
\end{equation}
where \trxpoll{i} is the RX timestamp of \pollpkt at \lrespi, and
\tprepare is the time needed to prepare the radio for transmission.
The term $i\tshift = \tshifti$ staggers the transmissions of the $n$
responders: each responder \lrespi adds a fixed, known delay
proportional to \tshift to the $\resppkt_i$ transmission, according to
the order defined by its index $i\,{\in}\,[0,n)$. The earliest
responder, \lrefanchor, adds no any delay and is referred to as the
\emph{main responder}.

The value of \tshift, compensated for by the tag in the subsequent CIR processing step (\cref{fig:scheme}), must be sufficiently large to accommodate the peak of a given responder \lrespi \emph{along with} its
MPCs; otherwise, the latter may ``spill'' into the CIR segment in which the peak of $\lresp_{i+1}$ is expected, potentially compromising its correct detection, as later illustrated in~\cref{fig:stscir_analysis}. 
In our environments~(\cref{sec:methodology}), $\tshift\seq{=}\SI{250}{ns}$, \ie $\sfrac{1}{4}$ of a preamble CIR symbol duration, is enough to discern the peaks of 
all 
responders and enable correct decoding.

\boldpar{Transmitting \resppkt packets.} 
The process above hinges on the ability of UWB radios to respond exactly at time \ttx{i}. 
However, while modern UWB radios can schedule transmissions, their TX time resolution is too coarse: 
$\SI{8}{\nano\s}$ for the Qorvo DW1000 and DW3000.
Given that $\SI{1}{\nano\s}$ corresponds to approximately $\SI{30}{cm}$ of propagation distance for radio waves,
an $\SI{8}{\nano\s}$ uncertainty in \ttx{i} would cause large errors in both CTX-based ranging and localization approaches.

Gro{\ss}windhager et al.~\cite{grosswindhager2019snaploc} circumvent
the problem by relying on an out-of-band infrastructure through which
the intended
timestamp is communicated by the anchors to the \tdoa
solver, which can therefore compute the position with 
full \ttx{i} resolution. 
Instead, Corbal{\'a}n et al.~\cite{corbalan2020ultra} mitigate the issue directly at the transmitter, by observing that the low TX time resolution is due to the Qorvo radios zeroing the last 9~bits of the intended TX timestamp \ttx{i}. 
Therefore, the error can be determined precisely at the transmitter, which can also compensate for it by adjusting (\emph{trimming}) the clock frequency while preparing the transmission of its \resppkt packet, a process known as \textit{clock detuning}. 
Essentially, a responder~\lrespi slows down its clock until the known error is eliminated, thus realigning the transmission time to the intended \ttx{i}.

\boldpar{CIR analysis.}  The two steps above ensure that the
concurrent \resppkt signals in the CIR are distinguishable and their
timing known with high temporal resolution.  The CIR analysis aims to
accurately derive the RX timestamps associated with each responder
\lrespi by executing the three following steps (\cref{fig:stscir}):

\noindent\circled{1} 
\emph{Responder association.} 
Based on the known scheduling of \resp packets, the CIR can be partitioned in \mbox{segments}, each intended to contain the peak of a responder~\lrespi. 
Ideally, their association with \lrespi follows the order determined by \cref{eq:ttx_rng}, starting with the main responder \lmainr. 
In practice, the CIR is stored in a \emph{cyclic} buffer, therefore the position~of~\lmainr is neither guaranteed to be first nor to be always the same. 
Corbal\'an~et~al.~\cite{corbalan2020ultra} tackle the issue by scheduling transmissions so that a peak-free CIR portion 
always exists between the peaks of the last responder $\lresp_{n-1}$
and \lmainr.  This empty portion
enables the correct identification of \lmainr, and, consequently, the association of each CIR segment to the correct responder. 

\noindent \circledb{2} \emph{First path detection.} 
The next goal is to detect the first peak contained in each segment, yielding a first, coarse estimate for the RX timestamp of the FP associated with the corresponding responder \lrespi. 
The simplest technique, adopted \eg by~\cite{grosswindhager2019snaploc}, determines the FP using a threshold \peakth on the CIR amplitude~(\cref{fig:stscir_analysis}). 
Another approach~\cite{corbalan2020ultra} relies instead on an adaptation of the 
\emph{Search \& Subtract} 
algorithm~\cite{dardari-toa-estimation}.

\noindent \circledb{3} \emph{First path refinement.} The previous step places the FP~in correspondence of a given CIR sample, whose time resolution ($1.0016\,\SI{}{\nano\s}$) is too coarse. 
The FP timestamp must hence be refined to achieve a sub-\SI{}{\nano\s} resolution: to this end, state-of-the-art approaches execute 
steps~1~and~2 
on a higher-resolution CIR, upsampled via Fast Fourier Transform (FFT)~\cite{corbalan2019chorus, grosswindhager2019snaploc, corbalan2020ultra}.


\vspace{-0.25mm}
\section{Challenges and Technical Approach} 
\label{sec:overview}
\vspace{-0.50mm}

\noindent
The state-of-the-art techniques illustrated in \cref{sec:ctxbasics}
have been instrumental in showing the \emph{feasibility} and benefits of CTX
for ranging and localization. 
However, 
they leave unaddressed several challenges severely hampering
practical applicability (\cref{sec:intro}). 
Next, we review these challenges, and provide a birds-eye-view of our technical approach
detailed later (\cref{sec:cloc-design}--\ref{sec:ondevice}). 

\vspace{-1.50mm}
\subsection{Challenges}
\label{sec:challenges}
\vspace{-0.50mm}

\newcounter{challengeCounter}
\newcommand{\challenge}[1]{
    \refstepcounter{challengeCounter}
    \label{challenge:#1}
}
\crefformat{challengeCounter}{#2C#1#3}
\Crefformat{challengeCounter}{#2C#1#3}
\newcommand{\challengePInternal}[2]{\boldpar{\cref*{challenge:#1}. #2}}
\newcommand{\challengePar}[2]{\challenge{#1}\challengePInternal{#1}{#2}}
\newcommand{\challengeRef}[1]{\cref{challenge:#1}}

\noindent A practical use of CTX-based techniques requires addressing three key needs, each posing a corresponding challenge. 

\begin{figure}[!t]
    \newlength{\figCtxCostsWidth}
    \setlength{\figCtxCostsWidth}{.975\columnwidth}
    \centering
    \vspace{-1.25mm}
    \includegraphics[width=\linewidth]{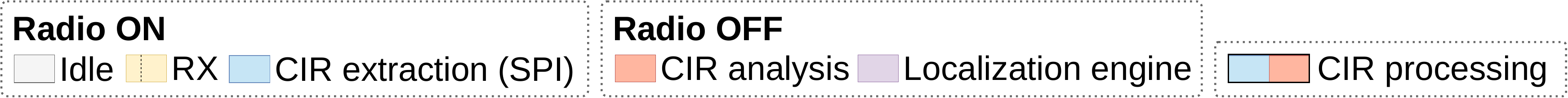}\\
    \vspace{+0.5mm}
    \begin{subfigure}[t]{0.4757653061\figCtxCostsWidth} 
        \centering
        \includegraphics[width=\linewidth]{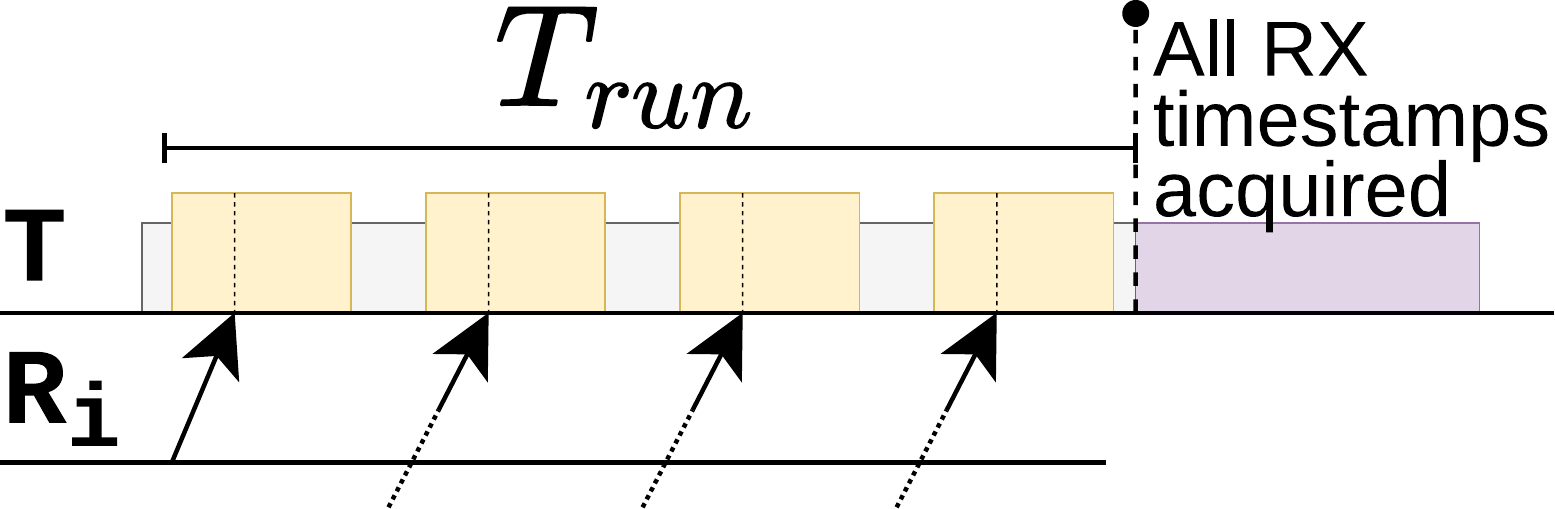}
        \vspace{-5.75mm}
        \caption{Traditional approaches}
        \label{fig:comp_staggered}
    \end{subfigure}%
    \hfill
    \begin{subfigure}[t]{ 0.5242346939\figCtxCostsWidth} 
    \centering
    \includegraphics[width=\linewidth]{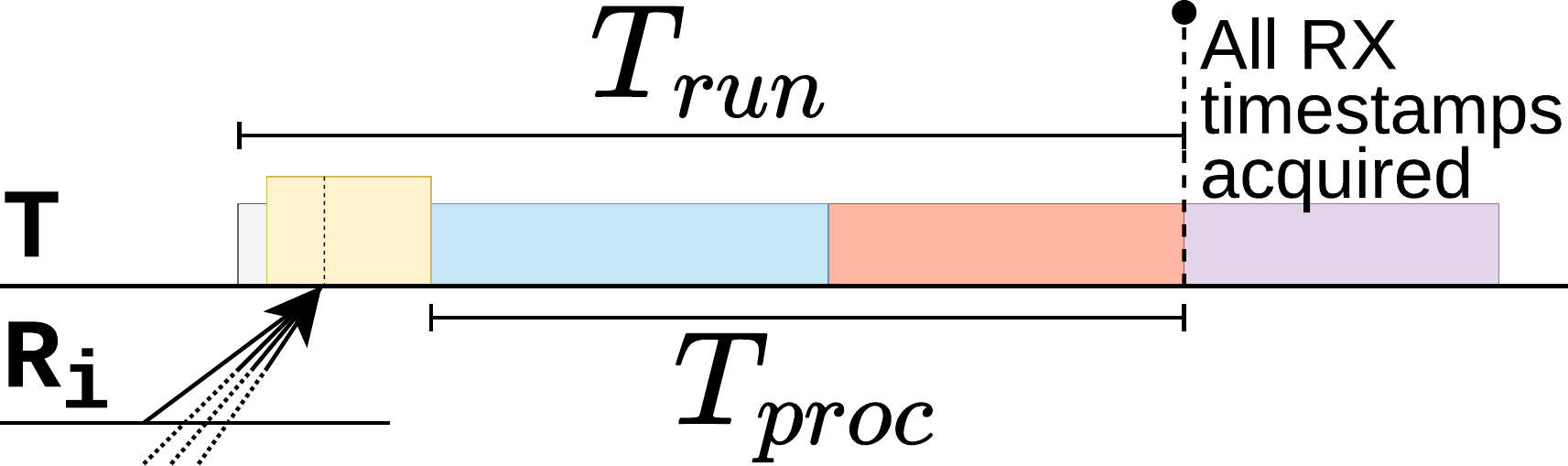}
    \vspace{-5.75mm}
    \caption{CTX-based approaches}
    \label{fig:comp_ctx}
    \end{subfigure}
    \vspace{-1mm}
    \caption{In contrast to traditional approaches, CTX-based ones require only a single frame reception, but entail processing the CIR in software before the RX timestamps can be derived. Additional processing is required in either cases for localization, \ie if deriving a position estimate is needed.} 
    \label{fig:ctx-costs}
    \vspace{-3.00mm}
\end{figure}

\challengePar{OnDevice}{The need for efficient on-device operation.}
\mbox{CTX-based} approaches 
combine the individual receptions at the tag of the sequential \resp
transmissions in staggered SS-TWR (\cref{fig:ss_twr_staggered}) and
\dltdoa (\cref{fig:dl_tdoa}) into a \emph{single} reception 
by forcing all transmissions to overlap in time.
This 
\emph{always} shortens the radio on-time, as only one frame reception
is necessary instead of multiple ones.  In principle, this enables
significant gains in energy consumption, channel occupancy, and update
rate.  In practice, CTX-based approaches require to process (\ie
extract and analyze) the CIR after reception; thus, for CTX-based
solutions to outperform non-concurrent ones, they must yield a shorter
\emph{overall} duration \truntime, from the start of reception to the
determination of all RX timestamps (\cref{fig:ctx-costs}).

Still, UWB radios do not support CTX processing natively, requiring software-based CIR extraction and analysis (§III). Existing CTX-based solutions~\cite{corbalan2018concurrent,
  grosswindhager2018concurrent, corbalan2020ultra, corbalan2019chorus, heydariaan19R3, 
  grosswindhager2019snaploc, heydariaan20anguloc, smaoui21single, tianyu20SingleAnchor,smaoui21anchor} did not address this constraint, as they rely entirely on \emph{offline} processing~on~powerful hardware, making them unsuitable for on-device operation.

\challengePar{LightRobustOp}{The need for lightweight \underline{and} robust operation.}
Efficient on-device operation is not the only reason to revisit CTX
techniques, as existing approaches \mbox{focused} on feasibility and
largely overlooked~\textit{robustness} and \emph{optimization
  opportunities}.

For instance, clock detuning~\cite{corbalan2020ultra}---the only
option for high-resolution scheduling without out-of-band
infrastructure---does \textit{not} exploit the full trimming range of
the radio and does \textit{not} account for clock drift, thus
operating slower than needed and leaving a potential error source
unaddressed.
Similarly, enforcing a large peak-free portion in~the
CIR~\cite{corbalan2020ultra} enables correct association of the first
peak in~each segment to its responder (\cref{fig:stscir},
step~\circledb{1}), but at the cost of reducing either
\begin{inparaenum}
\item the number of CIR segments, limiting the 
  number of supported responders, or
\item the segment length, limiting the maximum distance between
  devices and resilience to MPCs.
\end{inparaenum}

Another example is the Search\,\&\,Subtract algorithm used for FP~detection
in~\cite{corbalan2020ultra}, too computationally demanding for
on-device execution.  Alternative techniques
(step~\circledb{2}) 
rely on a single amplitude threshold
\peakth~\cite{grosswindhager2019snaploc,corbalan2019chorus}, whose
determination involves a poorly understood trade-off between
sensitivity to weak signals and resilience to
MPCs.  
If \peakth is too high, weak responder signals may go undetected; if
too low, false positives may appear due to noise or MPCs.
\cref{fig:stscir_analysis} exemplifies the latter~case:~the signal
from $\lresp_1$ exhibits two extra peaks with
amplitude~$>$\,\peakth~due~to~MPCs.  The first one is~close~to the FP
and can be rejected geometrically;
instead, the second ``spills'' into the next CIR segment and, being
its first peak, is mistakenly associated
to~$\lresp_2$. 
A more robust FP~detection technique maintaining high sensitivity
to weak responder signals while suppressing MPC-induced false
positives is required.

Similarly, upsampling the entire CIR via FFT to achieve higher temporal resolution and refine the FP (step~\circledb{3}) 
may also be prohibitively expensive on-device. 
Unfortunately, \emph{no alternative exists in the CTX literature}: a new and robust one must be devised to enable on-device operations. 

\challengePar{SignalSelect}{The need for signal selectivity.}
Existing CTX-based solutions are \emph{all} designed under the
assumption that \textit{no unwanted signals are present during
  reception}.  However, in practice, transmission from other UWB
devices not involved in the CTX exchange may occur along with
it. These unwanted signals (\eg $U_0$ and $U_1$
in~\cref{fig:stscir_interfered}) may generate extra peaks with amplitude $>$\,\peakth near the
expected ones (from responders $\lresp_1$ and $\lresp_3$) or even overlap with
them, challenging FP detection. 
These unwanted signals may originate from devices external to the system, but also from those belonging to it. 
For instance, in CTX-based \dltdoa systems, \emph{all} \mbox{anchors} transmit continuously: in large areas with many anchors in range, the CIR obtained by a mobile tag can accommodate only a 
subset of them, with the others becoming interferers.

Unfortunately, existing approaches do not tackle these problems, as they are evaluated in a \emph{closed system} setup where the UWB devices are \emph{exactly} those expected to participate~in CTX exchanges~\cite{corbalan2018concurrent, grosswindhager2018concurrent, corbalan2020ultra, corbalan2019chorus, heydariaan19R3, grosswindhager2019snaploc, heydariaan20anguloc, smaoui21single, tianyu20SingleAnchor,smaoui21anchor}. 
Scalability to large areas and resilience to UWB interference, both crucial in real-world scenarios, therefore remain unaddressed. However, both can be solved by empowering a tag to \textit{selectively receive signals} only from a dynamically-chosen set of responders and ignore all others.


\vspace{-1.25mm}
\subsection{Technical approach}
\label{sec:approach}
\vspace{-0.50mm}

\noindent
The challenges above are intertwined and so are the techniques we
propose as a solution. \cref{fig:scheme} (left) shows at-a-glance their
relations and role in the overall CTX-based operations
(\cref{sec:ctxbasics}). At a high level, these techniques can be grouped along the following dimensions, mirrored by the paper structure.

\boldpar{Revisiting the fundamentals of CTX operations
  (\cref{sec:cloc-design}).} 
Challenges \challengeRef{OnDevice} and \challengeRef{LightRobustOp} demand a rethinking of core CTX techniques (\cref{sec:ctxbasics}). 
We redesign the clock detuning approach~for high-resolution scheduling presented in~\cite{corbalan2020ultra} to take advantage of the full radio trimming range and provide clock drift estimation (\cref{subsec:highsched}).  
We enhance
the robustness of~CIR~analysis~with a novel responder identification technique ensuring that, \emph{during packet detection}, the main
responder peak is placed at a known position in the CIR
buffer~(\cref{subsec:txid}), providing a reliable base for
associating the remaining peaks to the correct responders. 
We further ensure robust FP detection by replacing the
single-threshold technique with a \emph{two-stage~detector}~(\cref{subsec:fpdetect}), and propose different lightweight techniques for FP refinement~(\cref{subsec:fprefine}), ascertaining their robustness experimentally. 

\boldpar{Enabling signal selectivity (\cref{sec:concurrentsts}).}
While revisiting the core CTX techniques, we consider not only the
base \cirp from the preamble, but also the \cirs from the STS
(\cref{subsec:background_basics}), our foundation to tackle
challenge~\challengeRef{SignalSelect}.  In~\cirs, signals from
transmitters lacking the correct cryptographic key
are filtered out:~$U_0$~and~$U_1$~in~\cref{fig:stscir_interfered}
would be absent, yielding a signal like the uninterfered one in the
\cirp of~\cref{fig:stscir_analysis}.

Still, achieving via STS the selectivity implied in
\challengeRef{SignalSelect} requires other mechanisms.  For instance,
the \cirs buffer~is half the size of the \cirp one, limiting the
number~of responder peaks that can be accumulated. We exploit the
ability of the popular Qorvo DW3000 family to yield two CIRs for the
first and second half of the STS, and design a \emph{key switching}
technique to capture and distinguish signals from different sets of
responders \emph{with a single packet reception}.

\boldpar{Efficient on-device processing (\cref{sec:ondevice}).} To
fully address challenge~\challengeRef{OnDevice}, the techniques above must be complemented
by system-level ones that further reduce computing demands and execution time \truntime (\cref{fig:ctx-costs}) without
sacrificing accuracy. 

We focus on reducing the delays induced by both CIR pre-processing and
CIR extraction over SPI.  Specifically, since converting CIR samples
to amplitudes requires costly square and square-root operations, we
propose the use of an \mbox{approximate} \mbox{conversion} that avoids
them while preserving accuracy~(\cref{subsec:fastcir}).  Moreover, we
design techniques that do not require the \emph{entire} CIR and
parallelize its extraction via DMA, further reducing processing
time~(\cref{subsec:bounded}).


\vspace{-0.25mm}
\section{Experimental Methodology}
\label{sec:methodology}
\vspace{-0.25mm}

\begin{figure}[!t]
    \centering
    \vspace{-3.25mm}
    \includegraphics[width=0.575\columnwidth]{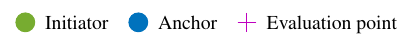}
    \\
    \vspace{-1.25mm}
    \begin{subfigure}[t]{0.475\columnwidth} 
        \centering
        \includegraphics[width=\linewidth]{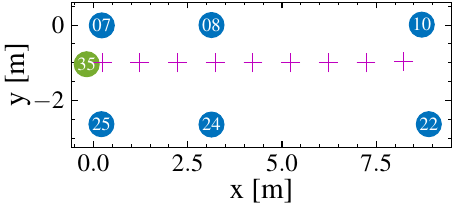}
        \vspace{-5.75mm}
        \caption{\envitihallway}
        \label{fig:tug_env}
    \end{subfigure}%
    \hfill
    \begin{subfigure}[t]{0.475\columnwidth} 
        \centering
\includegraphics[width=\linewidth]{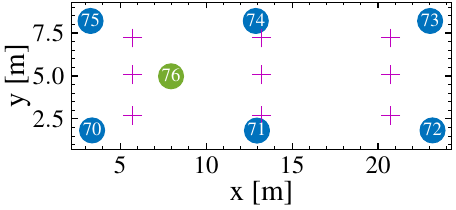}
        \vspace{-5.75mm}
        \caption{\envtuthallway}
        \label{fig:cloves_env}
    \end{subfigure}
    \vspace{-1.25mm}
    \caption{Setups used in our experiments. Circles indicate fixed responder nodes (blue) or the device serving as initiator in \dltdoa (green) and their IDs. 
    Magenta marks show the positions of the tag in the experiments presented in \cref{sec:cloc-design}--\cref{sec:concurrentsts}.}
    \label{fig:envs}
    \vspace{-2.75mm}
\end{figure}

\noindent 
In the rest of the paper, we present our techniques hand-in-hand with
their experimental evaluation
(\cref{sec:cloc-design}--\ref{sec:ondevice}), quantitatively
investigating design tradeoffs and improvements over the state of the
art. 
This technique-focused perspective is later complemented by a
system-wide analysis (\cref{sec:putting_things_together}) where all
techniques are combined into \cloc, a CTX-based \dltdoa system
whose 
performance demonstrates the validity of our approach.  Here, we
provide the appropriate context by offering salient details of our
experimental methodology.

\boldpar{Hardware and firmware.}  We base our results on the Qorvo
DWM3001CDK platform~\cite{qorvoDW3001cdkPage} hosting the DWM3001C
module~\cite{qorvoDWM3001cPage}, which embeds a DW3110
radio~\cite{qorvoDW31110datasheet}~and~a~nRF52833 SoC~\cite{nrf}.
This choice is a point of departure from all CTX-based systems to
date, which are based on the much older DW1000 radio family.  The
crucial difference is that DW3000 radios
support the STS, 
whose \cirs is stored~in two accumulators of 512~samples each.
Moreover, \cirp and \cirs samples are \SI{6}{\byte} \vs \SI{4}{\byte}
in~the DW1000, yet the transfer of a 
1016-sample \cirp is 
quicker in the DW3110 \mbox{($1.5$ \vs \SI{1.6}{\milli\s})} due to its
faster SPI \mbox{($32$ \vs \SI{20}{\mega\hertz})}.
We configure the \dwradio with channel~5 (center frequency
  6489.6\,MHz, bandwidth 499.2\,MHz), preamble code~9, and
64~preamble symbol repetitions, 
yielding a preamble of $\SI{65}{\micro\s}$ and a pulse repetition
frequency (PRF)
of \SI{64}{\mega\hertz}.

\boldpar{Environment.} 
We perform most of our experiments~in~two \emph{real-world testbeds} with different characteristics~(\cref{fig:envs}). This allows us to assess the robustness and applicability
of our techniques across different environments. 
The first testbed (\tug, \cref{fig:tug_env}) is a local UWB installation at Graz University of Technology, in a hallway $\SI{10}{\m}$ in height. 
We~use 7~anchors, installed on a metal rail at \SI{2.7}{\m} height and covering an area of $\SI{10}\times\,\SI{3}{\m^2}$. 
Tags are installed on top of metallic tripods at \SI{2.7}{\m} height.
The second testbed (\cloves, \cref{fig:cloves_env}) is the public 
UWB facility~\cite{molteni2022cloves,clovesurl} at the \envtut. 
\cloves nodes are
installed on the ceiling at $\SI{2.7}{\m}$ height; again, we use 7~of them as anchors, covering an area of $\SI{20}\times\SI{7}{\m^2}$. Tags are mounted on wooden poles at \SI{1.5}{m} height.


\vspace{-0.25mm}
\section{Revisiting the Core CTX Techniques}
\label{sec:cloc-design}
\vspace{-0.25mm}

\noindent
The first dimension of our approach~(\cref{sec:approach}) involves~revisiting state-of-the-art techniques to fulfill the need for efficient and robust on-device operation (challenges \challengeRef{OnDevice}--\challengeRef{LightRobustOp} in \cref{sec:challenges}).


   \subsection{High-resolution scheduling}
\label{subsec:highsched}

\noindent 
After receiving a \pollpkt packet, concurrent responders need~to transmit a \resppkt packet \emph{exactly} at the time specified by~\cref{eq:ttx_rng}.

\begin{figure}[!t]
    \centering
    \vspace{-2.50mm}
    \begin{subfigure}[t]{1.0\columnwidth}
        \centering
        \includegraphics[width=\linewidth]{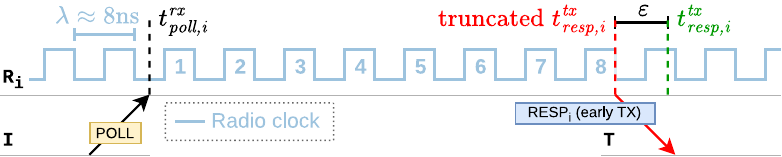}
        \caption{Without clock detuning. In this example, due to low TX scheduling resolution, responder $R_i$ anticipates the $\resppkt_i$ transmission (red \vs green vertical dashed lines), increasing ranging/localization error.}
        \label{fig:detexpl-wo}
    \end{subfigure}
    \hfill
    \vspace{+0.75mm}
    \begin{subfigure}[t]{1.0\columnwidth}
        \centering
        \includegraphics[width=\linewidth]{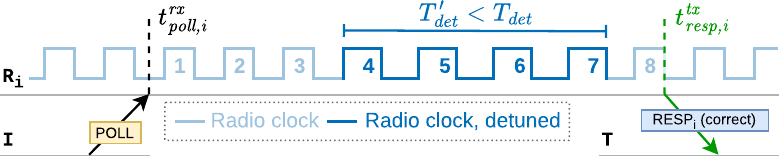} 
        \caption{With clock detuning. In this example, the radio clock is slowed down so that the $8^{th}$ radio tick aligns 
        with the intended TX time.}
        \label{fig:detexpl-with}
    \end{subfigure}
    \vspace{-0.5mm}
    \caption{Increasing TX resolution with clock detuning.}
    \label{fig:detexpl}
    \vspace{-2.75mm}
\end{figure}

\challengepar{Low TX resolution}
Unfortunately, both DW1000 and DW3000 radio families have the same low
TX~\mbox{scheduling} resolution of
\mbox{$\schedres \approx 1 / (\SI{125}{\mega\hertz}) \approx
  \SI{8}{\ns}$}, and truncate timestamps accordingly.  At a responder
\lrespi (\cref{fig:detexpl-wo}), this may result in an \emph{early}
transmission at \mbox{$\ttx{i}-(\ttx{i} \bmod \schedres)$}.
Since the receiving tag is unaware of the truncation, each \trx{i} value it extracts from the CIR is affected by a random error \mbox{$\schederr \in (-8, 0]~\SI{}{\nano\s}$}, potentially causing meter-level errors. 

\baselinepar{Slowing down the clock}
A finer TX resolution can be achieved by changing the responder's crystal oscillator frequency \fcbase to a lower one \fcnew~\cite{corbalan2020ultra}, in a process called \emph{clock detuning}~(\cref{sec:ctxbasics}). 
The frequency offset ratio \mbox{$\fcoff{=}(\fcnew{-}\fcbase)/\fcbase$},
when applied over a detuning interval \tdetune{}, corresponds to a time delay 
\mbox{$\schedcorr = -\tdetune{}\fcoff\approx-\schederr$},
realigning the TX timestamp to the intended untruncated \ttx{i}~(\cref{fig:detexpl-with}). 
Unfortunately, \fcoff cannot be set arbitrarily: only a limited set of
clock frequencies is available in the radio, selected via the
corresponding \emph{trim code}~(\cref{sec:challenges}, \challengeRef{LightRobustOp}). Thus, for
a given \tdetune, a \fcnew yielding exactly the desired \fcoff is
unlikely to exist. The baseline technique~\cite{corbalan2020ultra}
\begin{inparaenum}
\item picks the next available \fcnew that slightly \emph{exceeds} the
  desired correction, and
\item compensates the over-correction by decreasing the detuning interval to 
  $\tdetunep{}\seq{<}\tdetune{}$ yielding
  $\ronew=-\tdetunep{}\fcoff\approx-\schederr$.
\end{inparaenum}

\begin{figure}[!t]
    \vspace{-3.00mm}
    \centering
    \includegraphics[width=0.90\columnwidth]{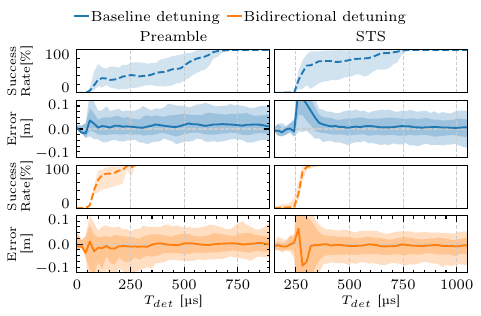}
    \vspace{-1.5mm}
    \caption{Reliability (success rate) and accuracy~(ranging error) of \sstwr 
    using different forms of detuning. RX timestamps are estimated from the preamble (left) or the STS (right).~For accuracy, the solid~line indicates the median; dark and light shaded regions represent the 90$^{th}$ and 99$^{th}$ percentile ranges. For reliability, the dotted line indicates the average success rate; the area shows the performance of best/worst responders.}
    \vspace{-2.75mm}
    \label{fig:schedErr}
\end{figure}

\methodpar{Bidirectional detuning} 
The detuning interval \tdetune is contained within the
  interval \tresp between the reception of \poll and the transmission
  of \resp. Therefore, \tdetune must be \emph{as small as possible},
  to correspondingly allow a small \tresp minimizing latency and
  energy consumption. In practice, this goal is challenging when the
  necessary compensation is as high as \SI{8}{\nano\s} and must be
  achieved with the limited range of frequencies \fcnew available on
  the radio, ultimately leading to increase \tdetunep and thus its
  upper bound \tdetune.

Still, we observe that the baseline is intrinsically limited by the
fact that it \emph{always} compensates starting from the truncated TX
time, yielding $\schederr\seq{\in}(-8, 0]~\SI{}{\nano\s}$.  An
alternative exists to reduce the maximum necessary compensation: to
first \emph{round} the ideal TX time to the closest radio tick,
yielding $\schederr\seq{\in}(-4, +4]~\SI{}{\nano\s}$, and only then
adjust the frequency accordingly---slowing down the clock when
$\schederr\seq{<}0$ (as in the baseline), or \emph{accelerating} it
when $\schederr\seq{>}0$.
The requirement to choose a new \fcnew that
  \begin{inparaenum}
  \item belongs to the set supported by the radio, and
  \item yields a valid compensation that can be completed within the
    original \tdetune,
\end{inparaenum}
still holds. However, our technique \emph{doubles} the options
available, since the viability of a given \fcnew can now be
ascertained not only when slowing down the clock, but also when
accelerating it.

\evalpar
We determine via \sstwr exchanges the \emph{minimum} \tdetune that
can be \emph{reliably} applied in the baseline detuning and ours, thus
determining which one allows for the shortest \tresp. We analyze:
\begin{inparaenum}
\item the \emph{success rate} of ranging exchanges, \ie the number of successful attempts over the total number of performed \sstwr exchanges, and 
\item the \emph{error} of successful ranging
  exchanges. Ranging failures can occur if \tdetune is so
    small that no available \fcnew can satisfy
    $\tdetunep\seq{<}\tdetune$.
Still, a successful \sstwr may
  nonetheless yield 
  unacceptably large ranging errors; both metrics are hence crucial.
\end{inparaenum}

Since the impact of clock detuning does not depend on the propagation
environment, we perform our evaluation only in \cloves
(\cref{fig:cloves_env}), using node 74 as initiator.  For each value
of \tdetune{} in~\cref{fig:schedErr}, we collect 300 \sstwr
measurements across all combinations of technique 
(baseline \vs bidirectional), responders (nodes~\mbox{70--73,~75}), and
RX timestamp estimation techniques (based on preamble \vs STS).
Responders do not transmit concurrently, allowing us to study the
impact of clock detuning without the risk of biasing results due to
CTX. However, unlike standard \sstwr, responders do \emph{not} embed
the TX timestamp in the \resp message
(\cref{subsec:background_schemes}), but mitigate the TX uncertainty
solely via clock detuning, as in real CTX
systems~(\cref{sec:ctxbasics}). Instead, we separately use standard
\sstwr without detuning as the ground truth for the ranging error.

\cref{fig:schedErr} shows the results. 
The advantage of our technique is evident by comparing the success rate of \sstwr exchanges.
Bidirectional detuning works reliably (\ie with no failures) with $\tdetune{}\geq\SI{348}{\us}$ for preamble-based ranging and
$\tdetune{}\geq\SI{396}{\us}$ for the STS;  
failures increase as smaller \tdetune{} values are used.
In
contrast, the baseline detuning requires a minimum \tdetune{} nearly
\emph{3~times longer}.
Notably, the gain attained by our technique is obtained \emph{without} detriment to ranging performance: bidirectional detuning has nearly zero median error \wrt the ground truth obtained without detuning.

\begin{figure}[!t]
    \centering
    \vspace{-3.25mm}
    \includegraphics[width=0.86\columnwidth]{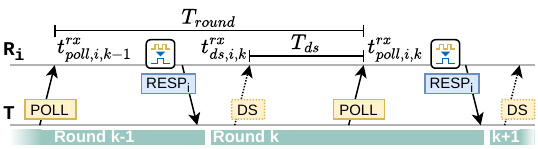}
        \vspace{-.5mm}
    \caption{The use of a double synchronization (\doublesync) allows to account for clock drift despite the use of clock detuning: \tds is not affected by detuning, unlike \tround.} 
    \label{fig:driftmethods}
    \vspace{-3.00mm}
  \end{figure}

\boldpar{Accounting for clock drift.} 
In the time interval between the reception of a \pollpkt packet \trxpoll{i} and the transmission of a $\resppkt_i$ packet \ttx{i}, the clocks of the various responders may drift apart, potentially causing a loss of synchronization. 
Each responder \lrespi must hence correct its drift before
transmitting $\resppkt_i$ by aligning its clock to the initiator, \ie
the tag in ranging or the reference anchor in \dltdoa
(\cref{subsec:background_schemes}).

\snaploc~\cite{grosswindhager2019snaploc}, relies on two consecutive
\pollpkt packets separated by a known interval \tround, estimating the
drift \drift{i} of $\lrespi$ \wrt the reference in a given round $k$
as
\begin{equation}
\driftest{i, k} 
= \frac{\left(\trxpoll{i, k} - \trxpoll{i, k-1}\right) - \tround}{\tround}.
\label{eq:drift}
\end{equation}

Unfortunately, the same approach cannot be exploited in our context:
during \tround, our detuning procedure introduces an \emph{artificial}
drift alongside the ``natural'' one experienced by nodes, preventing
the estimation of the latter via \cref{eq:drift}.
We thus opt for a \emph{double synchronization}
(\doublesync) technique, relying on two packets transmitted in the same round during a \emph{detuning-free} interval: \pollpkt, as 
in~\cite{grosswindhager2019snaploc}, and \dspkt, a second packet sent by the same source at an interval \tds before \pollpkt (\cref{fig:driftmethods}). 
Since during \tds no detuning is performed, each responder \lrespi can
estimate its natural drift as
\begin{equation}
  \driftest{i, k} 
  = \Bigl( \bigl(\trxpoll{i, k} - \trxds{i, k} \bigr) - \tds \Bigr) \,\, / \,\, \tds
\label{eq:driftds}
\end{equation}
and account for its clock drift by changing
\cref{eq:ttx_rng}~to
\begin{equation}
\ttx{i,k} = \trxpoll{i,k} + (1 + \driftest{i,k})(\tprepare + i\tshift).
\end{equation}

This approach ensures that responders transmit
their \resp packets based on a \emph{shared reference clock}, instead of their own,
yet does not require any change to the
detuning procedure.


   \subsection{Main responder identification} 
\label{subsec:txid}

\noindent 
In non-concurrent message schemes, tags can easily associate the RX timestamp of a \resppkt packet with its corresponding responder using the address encoded in the payload~(\cref{subsec:background_schemes}). 
In CTX-based systems, however, \emph{only one} payload can be decoded from multiple concurrent \resppkt messages, making timestamp-to-responder association more challenging and dependent solely on information extracted from the CIR~(\cref{sec:ctxbasics}).

\challengepar{Unpredictable responder ordering}%
In principle, the responder order predetermined via~\cref{eq:ttx_rng}
solves the problem. In practice, the use of a cyclic CIR buffer means
that UWB radios do \emph{not} necessarily place the peak of the main
responder \lrefanchor as first in the
CIR, 
disrupting the expected order~(\cref{sec:ctxbasics}). 
The first step of CIR analysis (\circledb{1} in \cref{fig:stscir}) is
therefore to identify the position of the main responder
\lrefanchor.

\newlength{\figwidth}
\setlength{\figwidth}{0.9\columnwidth} 
\begin{figure*}[!t]
    \centering    
    \vspace{-1.50mm}
    \begin{subfigure}[t]{1.25\figwidth}
        \centering
        \vspace{-32mm}
        \includegraphics[width=\linewidth]{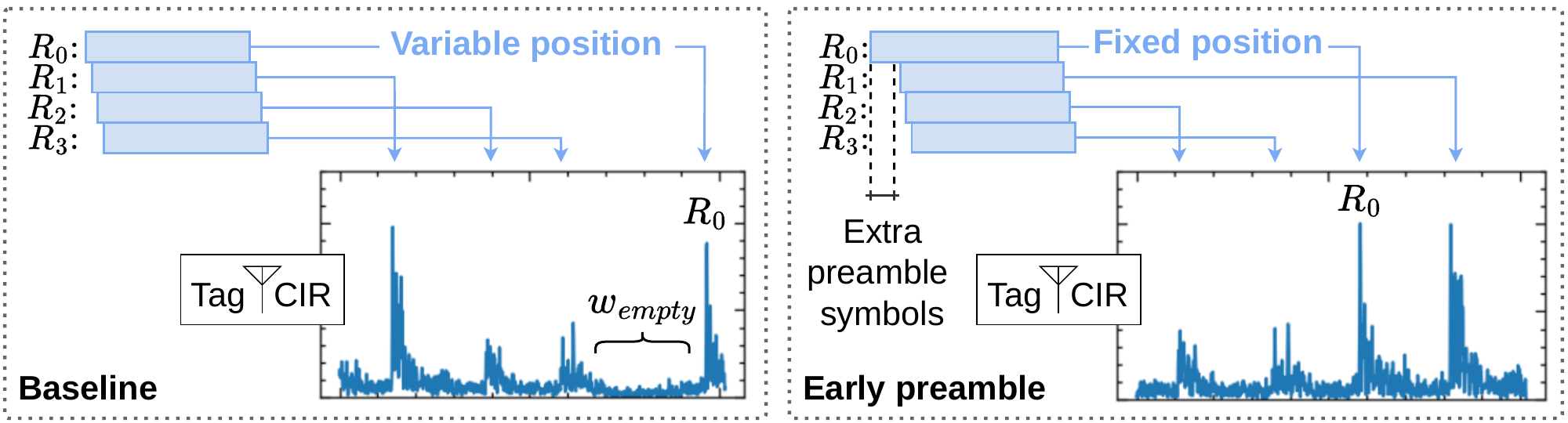}
        \vspace{-4mm}
        \caption{Main responder identification techniques}
        \label{fig:fpdet-methods}
    \end{subfigure}
    \hfill
    \begin{subfigure}[t]{0.40\figwidth}
        \centering
        \includegraphics[width=\linewidth]{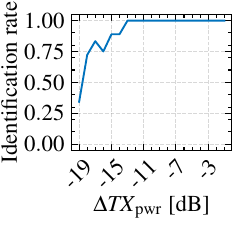}
        \vspace{-9mm}
        \caption{Baseline}
        \label{fig:fpdet-baseline}
    \end{subfigure}
    \hfill
    \begin{subfigure}[t]{0.302\figwidth}
        \centering
        \includegraphics[width=\linewidth]{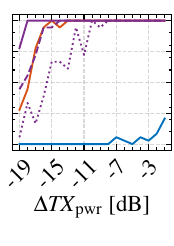}
        \vspace{-9mm}
        \caption{Early preamble}
        \label{fig:fpdet-earlyp}
    \end{subfigure}
    \hspace{1mm}%
    \begin{subfigure}[t]{0.15\figwidth}
        \centering
        \vspace{-34mm}
        \includegraphics[width=\linewidth]{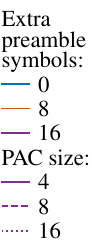}
    \end{subfigure}
     \vspace{-1.50mm}
     \caption{Illustration (a) and performance (b--c) of techniques
       for identifying the main responder \lmainr.  In the baseline
       approach, a portion of the CIR (\wnoise) must be kept
       peak-free. In the early preamble technique we propose,
       \lmainr transmits extra preamble symbols prior to the standard
       preamble sequence, which allows to predict the position in the
       CIR of the FP of \lmainr.}
    \label{fig:cfg-cir_reordering-pwr_vs_succrate}
    \vspace{-3.25mm}
\end{figure*}

\baselinepar{Empty CIR portion} To tackle this issue, Corbal{\'a}n et
al.~\cite{corbalan2020ultra} propose a technique scheduling \resppkt
transmissions such that a peak-free portion \wnoise is always present
after the segment allocated to the last responder; \lrefanchor can
then be easily identified by searching for
\wnoise. 
This technique, however, has two key drawbacks:
\begin{inparaenum}
    \item searching for \wnoise is expensive, as it requires to read and analyze \emph{the whole CIR}, and 
    \item for the technique to work reliably, \wnoise must be
      relatively large, 
      reducing the useful portion of the CIR and requiring a reduction
      of $\tshift$ in~\cref{eq:ttx_rng}, which in turn limits the
      supported area size and increases susceptibility to
      multipath~(\cref{sec:challenges},
      \challengeRef{LightRobustOp}). 
\end{inparaenum}

\methodpar{Early preamble} 
We propose an \mbox{alternative} technique to identify the FP peak position associated with~\lrefanchor within the CIR, which exploits the ability of the DW3000~radio to customize PHY settings and allow the transmission of extra preamble symbols \emph{before} the standard preamble sequence (\cref{fig:fpdet-methods}).
When applied \emph{only} to \lmainr, this enables the tag's radio to \emph{always} lock onto the 
early symbols~of~\lmainr, reliably placing its FP peak near index $750$, \ie where the (only) responder peak in non-concurrent schemes is typically placed. 

\configpar
We compare our technique against the baseline in terms of its ability to correctly identify \lmainr. 
We define the \emph{identification rate} as the fraction of attempts that correctly identify \lmainr. 
An attempt is considered successful when the tag estimates its distance with an error below \SI{0.5}{\meter}; larger errors suggest that a peak other than \lmainr was erroneously selected.

We ascertain the identification rate via micro-benchmarks
in an office environment with four~responders positioned at~$\SI{3}{\meter}$ from the tag 
concurrently replying to its \pollpkt packet. We systematically sweep the combination of three parameters:
\begin{inparaenum}
\item the \emph{difference in TX power} between \lmainr and the other~concurrent responders (\dtxpw), 
\item the \emph{number of extra preamble symbols} sent by \lmainr, and 
\item the \emph{PAC~size}, \ie the number of preamble symbols used for packet detection~(\cref{subsec:background_basics}). 
\end{inparaenum}
After identifying the CIR segment (presumably) corresponding to \lmainr, we acquire a coarse RX timestamp---needed to derive the distance between \lmainr and the tag---by leveraging the state-of-the-art FP detection technique proposed in~\cite{grosswindhager2019snaploc}.

Results show that the empty CIR portion left by~the baseline reliably
identifies \lmainr, but performance degrades at
\mbox{$\dtxpw\seq{\leq}{\SI{-13}{\decibel}}$}
(\cref{fig:fpdet-baseline}). Instead, our technique achieves perfect
identification down to~\SI{-15}{\decibel} with only 8~extra preamble
symbols 
and~\SI{-18}{\decibel} with 16.
Decreasing the PAC size, as in~\cref{fig:fpdet-earlyp} for 16~extra
symbols, largely improves the identification rate because, together
with the symbol increase, it increases the probability that the tag
locks onto the signal from \lmainr \emph{before} any other responder
transmits.

To validate these findings, we performed dedicated experiments in
\cloves and \tug, positioning the tag in the 9~evaluation points
marked in~\cref{fig:envs}, and exploiting the 4~outermost anchors as
responders. Results show that the early preamble technique
always achieves a \emph{perfect} main responder identification rate
with $\mathit{PAC}{=}4$ and 16~extra symbols. We therefore use this
configuration in the remainder of this work.

Moreover, our technique is not only highly reliable but also
\emph{intrinsically faster} than the baseline.  First, it eliminates
the need to read the \emph{entire} CIR: as discussed
in~\cref{sec:ondevice},
this 
allows to significantly reduce the  duration \truntime
of the overall CTX procedure. 
Second, it avoids the need to run an algorithm to detect the empty CIR portion, which by itself takes \SI{374}{\micro\s}.


   \vspace{-0.50mm}
\subsection{First path detection}
\label{subsec:fpdetect}
\vspace{-0.25mm}

\noindent
Once the main responder \lmainr is identified and the CIR is partitioned into segments, the next step is to detect the FP~in each segment to obtain a first, coarse estimate for the RX timestamp of the corresponding responder \lrespi~(\cref{fig:stscir}, \circledb{2}).

\challengepar{Reliably detecting the FP peak} Reliable FP detection is
especially challenging when the direct-path component is weak, \ie
exhibits a low amplitude, making difficult to distinguish from noise
or residual MPCs originating from earlier responders.  The detector
must therefore achieve a careful balance between sensitivity and
robustness, as overly conservative thresholds may cause missed
detections, whereas overly sensitive ones may lead to false
detections.

\baselinepar{Single threshold} In CTX-based systems, FP detection
commonly relies on adaptations of the Search\,\&\,Subtract
algorithm~\cite{corbalan2020ultra} or on a \emph{single} threshold
\peakth on the CIR
amplitude~\cite{grosswindhager2019snaploc,corbalan2019chorus}: since
the former is too computation-heavy for on-device execution, we select
the latter as our baseline. As in the corresponding
  works, we use a threshold \mbox{$\peakth = m \sigman$}, where $m$
  is a tunable multiplier and $\sigman$ the standard deviation of the
  noise, which we compute over 32~CIR samples. We then determine the
  FP of a responder \lrespi within its CIR segment $\ciri$ by using
  the \emph{first} CIR sample \fpsample \emph{after} the signal
  amplitude exceeds \peakth as the reference for the RX timestamp.

\begin{figure}[t]
    \centering
    \vspace{-3.50mm}
    \begin{subfigure}[t]{1\columnwidth}
        \centering
        \includegraphics[width=0.85\columnwidth]{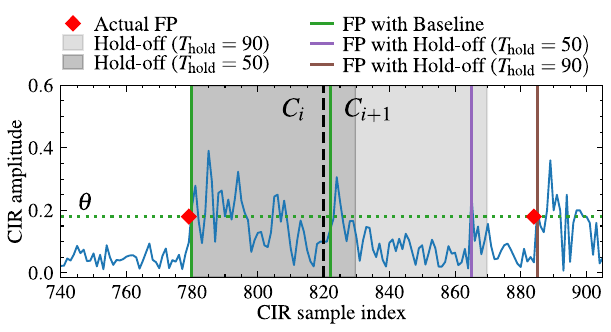}  
        \vspace{-1.75mm}
        \caption{Single threshold with and without hold-off interval:
          its use prevents the erroneous detection as FP of late MPCs
          extending into the following CIR segment.}
        \label{fig:fpdetection_illustration_single_stage}
            \vspace*{2mm}
    \end{subfigure}

    \begin{subfigure}[t]{1\columnwidth}
        \centering
         \includegraphics[width=0.85\columnwidth]{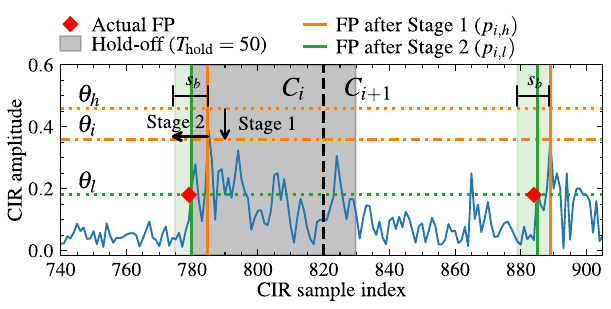}  
        \vspace{-1.75mm}
        \caption{Two-stage detector with hold-off: complementing
          amplitude-based detection with samples from nearby peaks
          within a search-back window $s_b$ further increases
          sensitivity to weak responder signals.}
        \label{fig:fpdetection_illustration_two_stage}
    \end{subfigure}
    \vspace{-1.00mm}
    \caption{FP detection techniques.
    }
    \label{fig:fpdetection_illustration}
    \vspace{-3.00mm}
\end{figure}

Unfortunately, as discussed
in~\cref{sec:challenges}~(\challengeRef{LightRobustOp}), a single
fixed threshold cannot simultaneously provide high sensitivity to weak
responder signals and robust suppression of false positives induced by
MPCs originated from earlier responders ``spilling'' into the next CIR
segment.  \cref{fig:fpdetection_illustration_single_stage} exemplifies
this: a single threshold \peakth allows correct FP detection in
segment \ciri (index~780), but detects as FP a late MPC extending into
segment \cirinext (index~822) instead of the actual FP (index~885).

\methodpars{Hold-off interval and two-stage detector} We propose two
techniques to improve the robustness of FP
detection: \begin{inparaenum} \item a \emph{hold-off interval} to
  explicitly avoid detecting MPC-induced peaks extending into
  subsequent CIR segments, and \item a \emph{two-stage detector}
  augmenting amplitude-based detection with the analysis of nearby
  peaks to distinguish responder signals from MPC-induced isolated
  peaks.\end{inparaenum}

\italicpar{Single threshold with hold-off interval.}  We extend the
baseline with a time interval during which all peaks
following a~FP~detection are ignored. After detecting a
FP in \ciri, the detector is paused for \thold samples, preventing
MPCs from the \emph{same} responder from being misidentified as a new
FP in \cirinext.
\cref{fig:fpdetection_illustration_single_stage} shows how pausing the
detector for $\thold\seq{=}90$ samples, until index~870, allows to
ignore MPC-induced isolated peaks and select the correct FP in
\cirinext (index~885), avoiding the false detection of the baseline.

Ideally, the hold-off interval is chosen to be as long as possible
while allowing FP detection to resume just before~the arrival of the
next FP.  If two consecutive responders $\lresp_i$ and $\lresp_{i+1}$
are equidistant from the tag, their FP arrivals are~separated by
$\delta$, as per \cref{eq:ttx_rng}, and the hold-off interval can be
maximized and set to $\thold=\delta$.  If, instead, $\lresp_i$ is
farther from the tag than $\lresp_{i+1}$ by a distance $\Delta d$, the
arrival of its response is delayed and the maximum hold-off interval
allowed is $\thold\seq{=}\delta\seq{-}\frac{2\Delta d}{c}$. Therefore, to
derive a \thold value valid for any tag~position within a given
environment, the worst-case difference in distance between two
responders must be determined. In our case~(\cref{fig:envs}), this is
$\Delta d_{\max} = \SI{9}{\meter}$ in \envitihallway, yielding
\mbox{$\thold = 190$} samples, and
\mbox{$\Delta d_{\max} = \SI{20.9}{\meter}$} in the larger
\envtuthallway, yielding a shorter \mbox{$\thold = 110$}~samples.

The effectiveness of the hold-off interval is limited in larger
environments, where the shorter \thold may still allow MPC-induced
false detections. This is illustrated in
\cref{fig:fpdetection_illustration_single_stage}, where a hold-off of
$\thold\seq{=}50$ samples still results in a MPC being falsely
detected as FP in \cirinext (sample index 865).

\italicpar{Two-stage detector with hold-off interval.}  To overcome
this limitation, we design a detector
exploiting the fact that a weak FP
  is often followed by stronger MPCs, recovering the genuine
  FP 
  when MPCs are misdetected.
Applied to each CIR \seg \ciri, the detector outputs the
    FP~index \fpsample and the corresponding threshold \peakthi, used
    as input to the subsequent FP~refinement
    (\cref{subsec:fprefine}), as follows
  (\cref{fig:fpdetection_illustration_two_stage}):
\begin{compactitem}
\item \emph{Stage 1: Adaptive high threshold}.  In \ciri, we search
  for the first sample exceeding a \emph{high} threshold
  \mbox{$\peakthi =\trhi=\mhi \,\sigman$}, storing its index
  \fpsampleh as the FP~candidate.  If none is found, $\mhi$ is reduced
  by $\zeta\seq{=}10\%$\ and the search restarts with the updated
  threshold \peakthi.
After $\eta\seq{=}5$ unsuccessful attempts, \ciri is marked empty and ignored, as further reductions increase false detections due to noise. 
$\zeta$ and $\eta$ were empirically determined in our two
  testbeds.
\item \emph{Stage 2: Auxiliary low threshold.} However, the
  FP~candidate above may be a strong MPC from the same responder,
  \ie an earlier, lower-power peak is the actual FP. To account for
  this possibility, we look in a small search-back window \sback for
  the earliest sample whose amplitude value equals or exceeds an
  additional \textit{low} threshold \mbox{$\trlo = \mlo \sigman$}.
If such a sample is found, we use its index \fpsamplel as the
  detected FP, $\fpsample\seq{=}\fpsamplel$, and correspondingly set
  $\peakthi\seq{=}\trlo$. Otherwise, the previous FP~candidate is
  confirmed, \mbox{$\fpsample\seq{=}\fpsampleh$} along with its
  threshold \peakthi.

By restricting the search with $\trlo$ to a small search-back window,
we increase the sensitivity of detection while minimizing the risk of
false positives, significantly higher if $\trlo$ were applied directly
to the entire \seg \ciri.
\end{compactitem}

Finally, the two-stage
detector \emph{still makes use of the hold-off interval}, \ie it
pauses for \thold samples after detecting a FP.

\boldpar{Parameter selection.}  FP detection depends on the multiplier
$m$, for the single-threshold detector, and on \mhi, \mlo, \sback, for
the two-stage detector.  For maximum performance, the values of these
parameters should be tailored to a specific deployment; still, their
tuning may be impractical in some real-world scenarios. Along
  these lines, we derive \emph{optimized} parameter configurations for
  \envtuthallway and \envitihallway. Moreover, we derive a single
  \emph{joint} configuration that, given the markedly different
  characteristics of our two setups, is likely directly usable as in
  new deployments, or can serve as a starting point to be refined with the
  same methodology we employ.

We perform 150~concurrent \sstwr rangings per tag position
(\cref{fig:envs}),
and process this dataset offline. We sweep all possible value
combinations and, for each, compute the \emph{detection rate} as
the fraction of samples where a FP is detected \textit{and} the
resulting ranging error is $<$\,\SI{0.5}{\meter}. 

\begin{figure}[!t]
    \centering
    \vspace{-3.5mm}
        \begin{subfigure}[t]{0.449\columnwidth}
          \centering
        \raisebox{-.15mm}{\includegraphics[width=\linewidth]{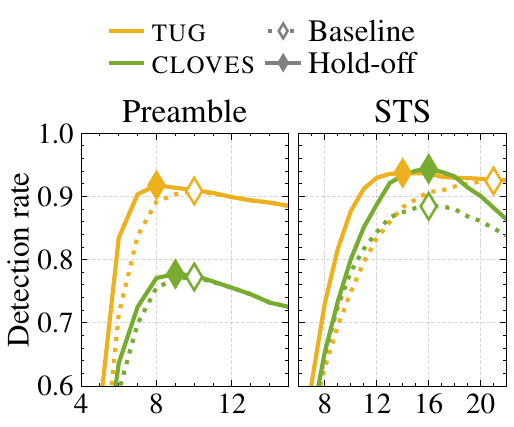}}
        \vspace{-5.5mm}
        \caption{Impact of $m$.}
        \label{fig:fp_parametrization_m}
    \end{subfigure}
    \hfill%
   \begin{subfigure}[t]{0.525\columnwidth}
    \centering
    \vspace{-31mm}%
        \includegraphics[width=1\columnwidth]{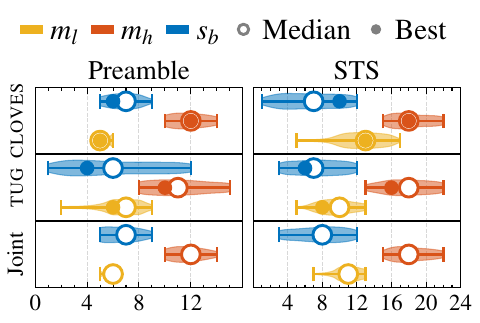}\\
    \hfill
        \vspace{-5.75mm}
        \caption{Range of values for \mlo, \mhi, \sback.} 
        \label{fig:cfg-fp_detection-fwbw-parameterspace}
    \end{subfigure}
    \vspace{-1.00mm}
    \caption{(a) Detection rate for different values of $m$, and (b)
      range (bars) and median (circle) of $\mlo$, $\mhi$, and $\sback$
      values within \perval{2.5} of the absolute best combination
      (dot).}
    \label{fig:fpnew_parametrization}
    \vspace{-2.75mm}
\end{figure}

We focus first on \cirp. For single-threshold detection,
$m\seq{=}10$ yields best performance in both setups
(\cref{fig:fp_parametrization_m}).
Still, in \envtuthallway the maximum detection rate is \perval{77},
confirming the intrinsic limitations of this baseline. Adding the
hold-off interval enables a lower threshold increasing detection
sensitivity resulting in lower best multipliers, $m\seq{=}8$
(\envitihallway) and $m\seq{=}9$ (\envtuthallway). Since, unlike the
baseline, these best $m$ values differ for our setups, hereafter we
determine the joint configuration applicable to both from the values
whose detection rate is within \perval{2.5} of the best one in
\emph{both} setups, taking their median (rounded down), yielding
$m\seq{=}9$.

For the two-stage detector, we focus on the combinations of
  \mlo, \mhi, \sback yielding a detection rate within \perval{2.5} of
  the maximum in each setup, and determine the value range for each
  parameter across these combinations
  (\cref{fig:cfg-fp_detection-fwbw-parameterspace}).
We derive the joint configuration from the combinations that are within
\perval{2.5} in \emph{both} setups, \ie the intersection of the
near-best combinations of \tug and \cloves, by taking the median
of each parameter over these combinations. This yields the values
$\mlo\seq{=}6$, $\mhi\seq{=}12$, and $\sback\seq{=}7$ used
hereafter. 

We repeated the procedure for \cirs, whose best values and
ranges differ from \cirp due to their slightly different characteristics~(\cref{subsec:pre-vs-sts}). The $m$ values we select for
\tug, \cloves, and joint are, respectively, $21,16,17$
for the baseline and $14,16,15$ with hold-off. For the two-stage
detector, the joint configuration is $\mlo\seq{=}11$, $\mhi\seq{=}18$, and
$\sback\seq{=}8$.

\begin{figure}[!t]
    \centering
    \vspace{-1.25mm}
    \includegraphics[width=1\columnwidth]{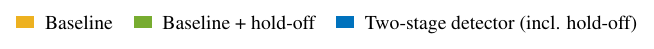} \vspace{-5mm} \\
    \begin{subfigure}[t]{0.3905\columnwidth}
        \centering
        \includegraphics[width=0.95\linewidth]{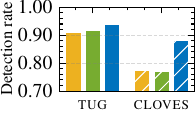}
        \vspace{-1.5mm}        
        \caption{\envitihallway--\cirp}
        \label{fig:fp_detection_pream_param_tug}
    \end{subfigure}
    \hfill
    \begin{subfigure}[t]{0.29\columnwidth}
        \centering
        \includegraphics[width=0.91\linewidth]{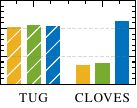}
        \vspace{-1.5mm}
        \caption{\envtuthallway--\cirp}
        \label{fig:fp_detection_pream_param_cloves}
    \end{subfigure}
    \hfill
    \begin{subfigure}[t]{0.29\columnwidth}
        \centering
        \includegraphics[width=0.91\linewidth]{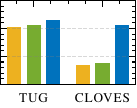}
        \vspace{-1.5mm}
        \caption{Joint--\cirp}
        \label{fig:fp_detection_pream_param_joint}
    \end{subfigure}
    \begin{subfigure}[t]{0.3905\columnwidth}
      \vspace{1mm}
        \centering
        \includegraphics[width=0.96\linewidth]{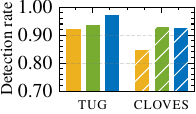}
        \vspace{-1.5mm}
        \caption{\envitihallway--\cirs}
        \label{fig:fp_detection_sts_param_tug}
    \end{subfigure}
    \hfill
    \begin{subfigure}[t]{0.29\columnwidth}
            \vspace{2mm}
        \centering
        \includegraphics[width=0.91\linewidth]{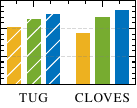}
        \vspace{-1.5mm}
        \caption{\envtuthallway--\cirs}
        \label{fig:fp_detection_sts_param_cloves}
    \end{subfigure}
    \hfill
    \begin{subfigure}[t]{0.29\columnwidth}
            \vspace{2mm}
        \centering
        \includegraphics[width=0.91\linewidth]{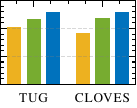}
        \vspace{-2.25mm}
        \caption{Joint--\cirs}
        \label{fig:fp_detection_sts_param_joint}
    \end{subfigure}
    \vspace{-1mm}
    \caption{
    Performance of the three FP detection techniques for different parameter configurations using \emph{preamble}-based CIRs (top) and STS-based CIRs (bottom). 
    Parameters are optimized for \envitihallway~(a, d), \envtuthallway (b, e), both environments~(c, f).
    }
    \label{fig:fpdetection_evaluation}
    \vspace{-3.50mm}
\end{figure}

\evalpar We compare techniques by detection rate, and assess a
setup-specific \vs a joint, generic
configuration. \cref{fig:fpdetection_evaluation} shows results in both
testbeds, separating the cases where techniques are optimized for \tug
(left), \cloves (center), or jointly (right); top charts use \cirp,
bottom ones \cirs.

  These results clearly show that, across all the combinations, the
  two-stage detector \emph{always} improves over the baseline---even
  when configured for one setup and applied in the other---and so does
  hold-off in all but one case (\cirp with \tug parameters in \cloves,
  \perval{-0.2}). Moreover, they also clearly show that the two-stage
  detector is the best choice, since in the only two cases where
  hold-off achieves a higher detection rate
  \begin{inparaenum}
  \item the difference is very small, \perval{0.3}~(\cref{fig:fp_detection_pream_param_cloves}) and~\perval{0.6}~(\cref{fig:fp_detection_sts_param_tug}),
  \item the two-stage detector still improves \wrt the baseline by \perval{0.3} and \perval{7.7}, respectively, and
  \item \cloc is configured with parameters optimized for the
    \emph{other} setup (\ie \tug with \cloves parameters
    in~\cref{fig:fp_detection_pream_param_cloves} and \cloves with \tug
    parameters in~\cref{fig:fp_detection_sts_param_tug}) rather than
    joint or specific ones.
  \end{inparaenum}
  Overall, with setup-specific parameters, the two-stage detector,
  which includes hold-off, improves from a minimum of \perval{3} in
  \tug to a remarkable \perval{15.5} in \cloves. In comparison, the
  improvement of hold-off applied to the baseline is appreciable only
  when using \cirs, up to \perval{5.9} in \cloves, while is very small
  (${\leq}0.9$\%) with \cirp.

Moreover, and equally important, the charts show that these
improvements can be seized also by the generic configuration catering
jointly for both setups, whose detection rate is the same or only
marginally lower (at most \perval{1.4}) \wrt the setup-specific
configuration yielding the absolute maximum. This is key because it
shows that, although a correct parameterization is important, the
sensitivity to differences \wrt the optimal value is very small, \ie
our technique is robust.

\begin{figure}[b]
    \centering
    \vspace{-5.50mm}
    \begin{subfigure}[b]{0.89\columnwidth}
        \centering
        \includegraphics[width=0.95\linewidth]{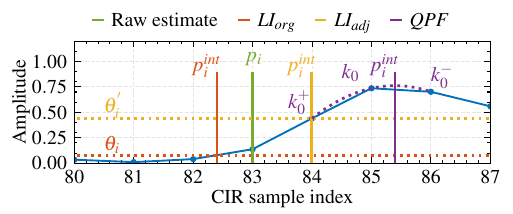}
         \vspace{-2.00mm}
        \caption{FP refinement techniques}
        \label{fig:cfg-fp_refinement-principle-illustration}
    \end{subfigure}\\
    \vspace*{1mm}
    \begin{subfigure}[t]{0.4086\columnwidth}
        \centering
        \includegraphics[width=\linewidth]{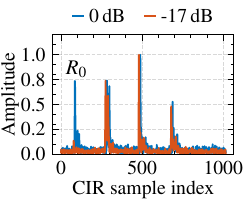}
         \vspace{-5.00mm}
        \caption{CIR}
        \label{subfig:cir}
    \end{subfigure}
    \hspace*{5mm}
    \begin{subfigure}[t]{0.3914\columnwidth}
        \centering
        \includegraphics[width=\linewidth]{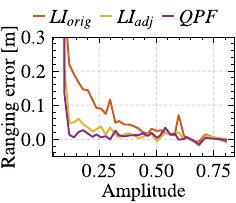}
        \vspace{-5.00mm}
        \caption{Ranging \vs amplitude}
        \label{subfig:fpref-rvsa}
    \end{subfigure}
    \vspace{-1mm}
    \caption{FP refinement. CIR amplitudes are normalized to the
      strongest peak.
      (\subref{fig:cfg-fp_refinement-principle-illustration})~Illustration
      of our techniques on a sample CIR. Horizontal lines are the
      thresholds in \liorg and \liadj; the purple curve is the
      polynomial output by QPF. (\subref{subfig:cir})~Impact on the
      CIR of the TX power difference between \lmainr and other
      concurrent responders. (\subref{subfig:fpref-rvsa})~Ranging
      error \vs amplitude of \lmainr.}
    \label{fig:cfg-fp_refinement-principle}
    \vspace{-1.25mm}
\end{figure}


   \subsection{First path refinement}
\label{subsec:fprefine}

\noindent 
The techniques in \cref{subsec:fpdetect} map the detected FP for each
responder \lrespi to a specific CIR sample index \peaki---in practice,
the first sample \emph{after} the signal amplitude crosses
the threshold \peakthi---from which the corresponding RX timestamp is derived.

\challengepar{Limited RX timestamp resolution} 
With this approach, however, the RX timestamp resolution is inherently limited by the CIR sampling interval ($\approx \SI{1}{\nano\s}$), which corresponds to a distance uncertainty of up to $\SI{30}{\cm}$. 
Therefore, a refinement technique is required to achieve \mbox{sub-ns} resolution and improve the coarse RX timestamp estimate (\circledb{3}, \cref{fig:stscir}).

\boldpar{No baseline with low computational constraints.}
State-of-the-art CTX-based systems improve the timing resolution by upsampling the entire CIR via FFT~\cite{corbalan2019chorus, grosswindhager2019snaploc, corbalan2020ultra}. 
However, this is too computationally demanding for on-device execution on off-the-shelf platforms such as the Qorvo DWM3001CDK. 

\methodpars{Linear interpolation and quadratic fitting} 
Since no existing approach can run on-device and serve as a baseline, we propose three lightweight refinement techniques (\cref{fig:cfg-fp_refinement-principle-illustration}) and evaluate their performance.  

\italicpar{Linear interpolation with original threshold (\liorg).}~Arguably, the simplest approach consists in identifying,~for~each FP index \peaki, the precise instant where the signal crosses the static threshold \peakthi used for FP detection. 
As illustrated~in \cref{fig:cfg-fp_refinement-principle-illustration}, this timestamp is calculated at the intersection~of~\peakthi and the line segment connecting \peaki and the preceding CIR sample ($\peaki - 1$). 
Denoting $\ciri[k]$ as the CIR amplitude~at~sample index $k$, the refined index \peakiint can thus be expressed as: 
\begin{equation}
 \peakiint = \frac{\peakthi - \ciri[\peaki-1]}{\ciri[\peaki] - \ciri[\peaki-1]} + (\peaki-1)
 \label{eq:interpolation}
\end{equation}

\italicpar{Linear interpolation with adjusted threshold (\liadj).} 
While simple to implement, a fixed threshold yields timestamps~that vary with the received signal power. 
Because high-power peaks exhibit steeper slopes, the line segment connecting \mbox{($\peaki-1$)} and \peaki would cross a fixed threshold~\peakthi earlier than with low-power peaks. 
To mitigate this power-dependent bias, we compute a different threshold \trad for each peak \peaki. 
First, we find the index \mbox{$\kmax \in [\peaki-1, \peaki + \peakidelta]$} with maximum amplitude $\ciri[\kmax]$, where $\peakidelta\seq{=}2$ was empirically selected to balance peak capture probability and search window size.
Then, we set the new threshold \trad to roughly half of the peak maximum amplitude, \ie \mbox{$\trad= \ciri[\peaki-1] + (\ciri[\kmax] - \ciri[\peaki-1])/2$}.
Finally, we compute 
\peakiint via Eq.\,\ref{eq:interpolation} by replacing \peakthi with \trad.

\italicpar{Quadratic pulse fitting (\qpf).} We propose another
technique mitigating power-dependent timing bias by directly
estimating the center of the signal pulse via parabolic interpolation,
\ie without a threshold. We first locate the first CIR sample
$\pfalling$\,$>$\,$\peaki$
exhibiting a negative gradient (\ie the first sample after the peak maximum) along with its two preceding neighbors, \mbox{$\pcenter=\pfalling-1$} and \mbox{$\prising = \pcenter - 1$} (\cref{fig:cfg-fp_refinement-principle-illustration}). 
We then model the pulse peak with a parabola \mbox{$y(x,\pcenter) = a(x - \pcenter)^2 + b(x - \pcenter) + c$} passing through the three points $(\prising, \ciri[\prising])$, $(\pcenter, \ciri[\pcenter])$, and $(\pfalling, \ciri[\pfalling])$. 
The refined timestamp is then set to the parabola vertex: \mbox{$\peakipul = \pcenter - (b /2a)$}. 
Compared to linear interpolation, this technique offers two key advantages: 
\begin{inparaenum} 
    \item by using the peak maximum and its adjacent samples, it operates in a high signal-to-noise ratio region, avoiding the noise-induced jitter typical of low-amplitude threshold crossings; and 
    \item parabolic interpolation of the peak vertex removes the need for threshold tuning, while remaining invariant to signal amplitude variations\end{inparaenum}.

    \evalpar We evaluate our techniques on the tabletop dataset
    from~\cref{subsec:txid}, where
    4~\mbox{responders} 
    concurrently reply to a tag \pollpkt and the TX~power~of the main
    responder \lmainr is varied to create a power difference
    \mbox{$\dtxpw\seq{\in}[-19, 0]$\,dB} \wrt the others.
    This mimics different distances of \lmainr and allows us
      to explore large differences in its peak amplitude, leaving the
      one of other responders essentially unchanged. \cref{subfig:cir}
      shows an example, with amplitudes normalized to the strongest
      peak to better highlight the relative differences.

    Ideally, ranging should be insensitive to variations in
      peak amplitude, unless it becomes too low.
      \cref{subfig:fpref-rvsa} shows, for each technique, the ranging
      error between the tag and \lmainr as a function of its
      (normalized) peak amplitude.  \liorg is the most affected by low
      amplitudes; errors begin increasing steadily already at $0.45$,
      due to its inability to adapt.
      \liadj and \qpf remove this limitation, and are markedly more
      robust; for amplitudes of $\seq{\approx}0.12$, errors are still
      below $4.8$ and $\SI{1.3}{\centi\meter}$, respectively,
      increasing dramatically only beyond this amplitude. Hereafter,
    we use \liadj: manual inspection of the CIRs showed its output
    timestamps are in closest agreement with the radio, and
    the improvement of QPF is marginal.


\vspace{+0.25mm}
\section{Enabling Signal Selectivity}
\label{sec:concurrentsts}
\vspace{-0.25mm}

\noindent
The techniques in \cref{sec:cloc-design} address challenges
\challengeRef{OnDevice}--\challengeRef{LightRobustOp}
(\cref{sec:challenges}) but, alone, cannot prevent CIR disruption due
to unwanted UWB signals (\challengeRef{SignalSelect}).  In principle,
the CIR of the STS (\cirs) could be processed with the same techniques
used for the preamble (\cirp), benefiting from the rejection of UWB
interference offered by STS.  In practice, this would reduce the
number of concurrent responders supported, due to the smaller size of
\cirs (512 \vs 1016 samples, \cref{sec:approach}).  Moreover,
  even if all responders would fit in \cirs, this would support only a
  small and closed system as with existing systems based on \cirp,
  preventing large-scale deployments where the set of nearby
  responders is selected dynamically (\challengeRef{SignalSelect})
  among the many available. We tackle both limitations via a novel
\emph{key switching} technique enabling the tag to accumulate not one
but two STS signals in separate \cirs, yet for the same reception.

\vspace{-0.75mm}
\subsection{Key switching}
\label{subsec:sts_keyswitching}
\vspace{-0.25mm}

\noindent
Our scheme relies on the availability of two distinct \cirs on
all radios in the \dwfamily.  These CIRs were originally
intended for Phase Difference of Arrival (PDoA) measurements in
dual-antenna devices, accumulating the same STS signal derived from
the same cryptographic key. Still, DW3000 radios \emph{always} return
two \cirs also in the common case when \emph{only one antenna} is
present, as in the \dwradio we use.

We exploit this general feature by assigning different keys \stsAKey
and \stsBKey to different sets of concurrently transmitting
responders, \eg $R_0$--$R_1$ and~$R_2$--$R_3$. Then, we
\textit{change} the STS key amidst
reception~\cite{stocker2022towards}, producing two \emph{distinct} CIR
estimates, \CIRone and \CIRtwo, correlating \emph{different} STS
sequences from \emph{distinct keys}, but derived from the \emph{same}
received frame. With \cirs holding two responders, a total of four
responders can be supported, as with the preamble-based techniques. We
detail next the techniques that make this possible.

\begin{figure}[!t]
\centering
\vspace{-2.00mm}
\includegraphics[width=0.825\columnwidth]{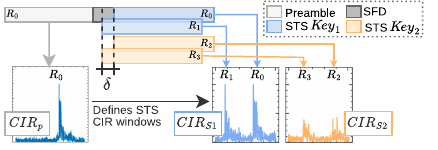}
\vspace{-2.00mm}
\caption{Concurrent STS message scheme. 
The receiver is configured to use \stsAKey for \CIRone and \stsBKey for \CIRtwo.}
\label{fig:cfg-keyswitching_message_scheme}
\vspace{-0.75mm} 
\end{figure}

\boldpar{New message scheme: Concurrent STS.} Nevertheless, responder
identification requires a different arrangement of concurrent \resppkt
packets~(\cref{fig:cfg-keyswitching_message_scheme}) \wrt the
preamble-based case~(\cref{fig:fpdet-methods}), as we can no longer
exploit the cyclic nature of \cirp.
In our scheme, the main responder \lmainr is the only node
transmitting a preamble;\footnote{A consequence is that $\lanchor_0$
  must be in the communication range with the tag; still, in large
  areas where this is not guaranteed, simple solutions exist albeit
  outside the scope of this paper. For instance, ranging initiators
  could include their targets in the \pollpkt packet starting with
  \lmainr, enabling dynamic selection of the main responder. In
  \dltdoa systems, anchors could be organized into cells, each
  containing a designated main responder.} since the relevant
concurrent signals are the STS ones, concurrent preambles would be a
useless complication. Therefore, the peak position for \lmainr is
known, always automatically placed by the radio close to the end of
\CIRone. As for~$\lanchor_1$, its peak can only fall in the empty
portion at the start of \CIRone, which we ensure at $\lanchor_1$ by
setting a \emph{negative} TX offset \tshift of the \resp transmissions
\wrt $\lanchor_0$. The other two responders are handled similarly: we
align the \resppkt of $\lanchor_2$ with $\lanchor_0$, causing the peak
of $\lanchor_2$ to appear at the end of \CIRtwo, and use the same
negative shift \tshift for $\lanchor_3$.
 
\boldpar{Hooking into symbol generation.}  The \dwradio
computes the expected sequence of STS symbols based on the configured
STS key, and stores them internally 
for correlation with the incoming signal. Key switching hooks
  into this procedure, replacing the key while symbols are being
generated; if timed correctly, some symbols are associated
with \stsAKey and the rest with \stsBKey.  To ensure proper
timing, the tag performs~key switching in two phases.  During
\textit{sync}, the MCU sets a fake reception, triggering an interrupt
used to synchronize the clocks of the \dwradio and the host MCU.
Then, during \textit{switch}, the MCU configures the \dwradio to
initiate reception, and schedules an SPI write command to change the
key.

\begin{figure}[!t]
    \centering
    \vspace{-3.25mm}
    \includegraphics[width=0.95\columnwidth]{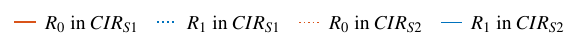}
    \\
    \vspace{-2.00mm}
    \begin{subfigure}[t]{0.62\columnwidth}
        \centering
        \includegraphics[width=0.44\linewidth]{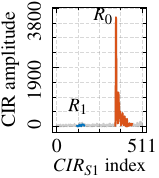}%
        \hspace{3mm}%
        \includegraphics[width=0.3\linewidth]{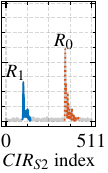}
        \vspace{-1.00mm}
        \caption{Impact of a large delay
          $\tkso\seq{=}\SI{140}{\micro\s}$.}
        \label{fig:keyswitch-wrongtau}
    \end{subfigure}%
    \begin{subfigure}[t]{0.31\columnwidth}
        \centering
        \includegraphics[width=\linewidth]{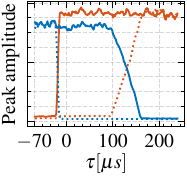} 
        \vspace{-5.25mm}
        \caption{\mbox{Amplitudes \vs \tkso.}}
        \label{fig:keyswitch-time}
    \end{subfigure}
   \vspace{-1.75mm}
    \caption{
      Choosing the key switching delay \tkso for two responders
      $\lanchor_0$ and $\lanchor_1$.
    }
    \label{fig:keyswitch}
    \vspace{-3.25mm}
\end{figure}

\boldpar{Timing characterization.} To avoid bias towards a particular
responder set, the tag must ensure that symbols from \stsAKey are
accumulated \emph{only} in \CIRone and likewise for \stsBKey and
\CIRtwo. We offer control over this aspect by performing key switching
with a tunable time delay \tkso \wrt the start of reception,
determining how STS symbols are split between
keys.  

We exemplify the impact of 
\tkso via a tabletop experiment with only two responders: $\lanchor_0$
using \stsAKey and $\lanchor_1$ using \stsBKey.  A proper \tkso
results in~a~single~responder~peak in \CIRone for $\lanchor_0$~and in
\CIRtwo for~$\lanchor_1$.  Instead, when \tkso is too large, key
switching occurs too late, causing the tag to accumulate in \CIRtwo
also STS symbols sent with \stsAKey, possibly producing an undesired
extra peak (\cref{fig:keyswitch-wrongtau}).
\cref{fig:keyswitch-time} offers a more comprehensive view, comparing
the peak amplitude of responders as a function of~\tkso both when
present in the intended \cirs (solid line) and in the \emph{wrong} one
(dashed).  We observe that \stsBKey should never be written earlier
than
$\tkso\seq{=}$$\SI{-18}{\micro\s}$; otherwise, the radio generates
\emph{all symbols} with it, causing the peak of
$\lanchor_1$ to appear in both CIRs and the signal from
$\lanchor_0$ to go undetected.  Moreover, the abrupt peak amplitude
change near this value shows that the first~64 of the 128~STS symbols
are \emph{pre-generated} 
from \stsAKey before reception.
Instead, the symbols from \stsBKey are generated \emph{on-the-fly},
since peak amplitudes change gradually in
$\tkso$\,$\in$\,$[96,160]$\,\SI{}{\micro\s}, \ie the second half of
the STS. 
In practice, when using an STS length of 128 symbols, the choice of
$\tau$ is rather flexible, with $\tau\seq{\in}[-10,90]$\,\SI{}{\micro\s}. In our setup, we set the STS length to 128 and $\tau=\SI{0}{\micro\second}$.

\vspace{-0.50mm}
\subsection{STS robustness and accuracy}
\label{subsec:pre-vs-sts}

\noindent 
To date, STS has never been exploited for CTX.  Therefore, we evaluate
the robustness and timestamping accuracy of our concurrent STS scheme
(\cref{subsec:sts_keyswitching}) compared to the classical preamble-based
one.
We perform experiments with and without UWB interference, adopting the
same experimental methodology in~\cref{sec:cloc-design}, including tag
positions~(\cref{fig:envs}).

\begin{figure}[!t]
    \centering
    \vspace{-2.75mm}
    \includegraphics[width=0.99\columnwidth]{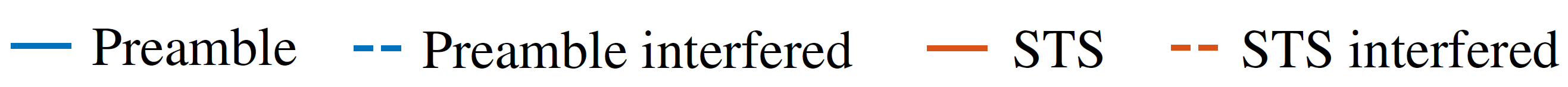}
    \\
    \vspace{-0.5mm}
    \begin{subfigure}[t]{0.4677\columnwidth}
        \centering
        \includegraphics[width=\linewidth]{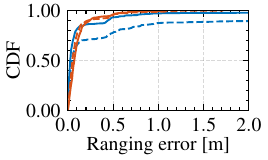}
        \vspace{-5.75mm}
        \caption{\envitihallway}
    \end{subfigure}
    \begin{subfigure}[t]{0.3793\columnwidth}
        \centering
        \includegraphics[width=\linewidth]{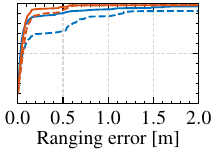}
        \vspace{-5.75mm}
        \caption{\envtuthallway}
    \end{subfigure}
    \vspace{-2.00mm}
    \caption{Absolute error of preamble- and STS-based concurrent ranging, with and without UWB interference.
    }
    \label{fig:STS-uwb_interference}
    \vspace{-2.25mm}
\end{figure}

\boldpar{No UWB interference.} The reliability of key switching and
overall viability of STS-based CTX are confirmed by an FP~detection
rate \perval{{\geq}98} in both testbeds, as with the preamble.  As for
ranging accuracy (\cref{fig:STS-uwb_interference}), the STS is less
prone to large errors than the
preamble.
In \tug, $98.1\%$ of rangings have error $<$$\SI{0.5}{\meter}$ with
STS, against $95.1\%$ with preamble; in \cloves, these values are
$96.7\%$ and $90.4\%$, respectively.
In neither case we observe significant differences across responders.

\boldpar{Under UWB interference.}  We also study whether concurrent
STS can effectively protect timestamps from UWB interference.  To this
end, we let two additional nodes (node 8 and 24 in \envitihallway,
71 and 74 in \envtuthallway) transmit data while 
\resppkt packets are sent by the intended responders, and assess the
impact of this unwanted interference on the ranging error.

Compared to the intended UWB signals, we use a different preamble code for the interfering preamble, and a different key for the interfering STS. 
We choose this approach for two reasons: 
\begin{inparaenum}
\item 
an interfering signal generated using the same STS key would imply that the interference source has access to our cryptographic secrets; and
\item when using the same preamble code, 
the tag's radio might lock onto the interfering signal rather than the intended preamble, resulting in misaligned peaks in the \cirs hindering the STS timestamping process.
\end{inparaenum} 
\emph{Neither case can be addressed by the STS, even in its
  non-concurrent version.}  Using different preamble codes and STS
keys is thus more informative, as it allows us to directly determine
the extent to which preamble-based CTX can withstand interference,
compared to STS-based CTX.
Moreover, some degree of UWB interference is also intrinsic in our key
switching approach, since the STS received by a tag and based on one key is interfered by the other still being transmitted. 

\begin{figure}[t]
    \centering
    \vspace{-1.75mm}
    \begin{subfigure}[t]{0.4332\columnwidth}
        \centering
        \includegraphics[width=\linewidth]{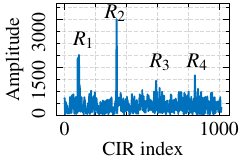}
        \vspace{-5.25mm}
        \caption{\cirp}
    \end{subfigure}
    \hfill
    \begin{subfigure}[t]{0.1811\columnwidth}
        \centering
        \includegraphics[width=\linewidth]{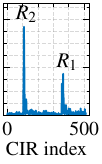}
        \vspace{-5.25mm}
        \caption{\CIRone}
    \end{subfigure}
    \hfill
    \begin{subfigure}[t]{0.1817\columnwidth}
        \centering
        \includegraphics[width=\linewidth]{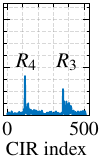}
        \vspace{-5.25mm}
        \caption{\CIRtwo}
    \end{subfigure}
    \vspace{-1.25mm}
    \caption{Impact of UWB interference on preamble- and STS-based
      CTX. The cross-correlation noise in the \cirp makes FP detection
      for $\lanchor_2$ and $\lanchor_3$ hard if not
      impossible. Instead, responder peaks are clearly discernible in
      the two \cirs. 
    }
\label{fig:cfg-uwb-interference_preamble_vs_sts}
    \vspace{-3.00mm}
\end{figure}

We find that, when using the preamble under UWB interference, the FP
detection rate decreases by $11\%$ in \envitihallway and
\envtuthallway, whereas it is \emph{unaltered} when
using the STS. The reason is the increased robustness to noise of the
STS signal, exemplified
in~\cref{fig:cfg-uwb-interference_preamble_vs_sts}, 
which is reflected also on the ranging error, whose distribution we
report in~\cref{fig:STS-uwb_interference} for samples with successful
FP~detection.  STS-based concurrent ranging is marginally affected by
UWB
interference: 
in both testbeds, the median error increases only by
\SI{2}{\centi\meter} \wrt the case without interference. In contrast,
the accuracy of the preamble-based version degrades significantly,
with errors ${>}\SI{0.5}{\meter}$ in \perval{19} and \perval{28} of
the cases in \envitihallway and \envtuthallway,
respectively.

\section{Enabling efficient on-device processing}
\label{sec:ondevice}

\noindent 
The techniques in~\cref{sec:cloc-design}--\ref{sec:concurrentsts}
bring large improvements over the state of the art, both in terms of
performance and functionality.  However, to seize their full
potential, they must be complemented by an \emph{efficient on-device
  implementation}, minimizing computing demands and hence the overall
execution time and energy consumption.  We focus on the pre-processing
of the CIR~(\cref{subsec:fastcir}) and its extraction over
SPI~(\cref{subsec:bounded}).

\subsection{Fast calculation of CIR amplitudes}
\label{subsec:fastcir}

\boldpar{Exact amplitudes.} 
Existing CTX systems assume that the CIR is transferred to a more powerful device for \emph{offline} pre-processing and analysis~(\cref{sec:challenges}, \challengeRef{OnDevice}). 
The mismatch between this approach and \emph{on-device} operation is already evident in the first pre-processing step: \emph{calculating the amplitude} of each CIR sample, which serves as input to FP detection and refinement (\cref{subsec:fpdetect}, \cref{subsec:fprefine}). 
CIR samples are complex numbers $(r,i)$, whose 
\emph{exact} amplitude, \ie $\sqrt{r^2+i^2}$, requires \emph{square and square-root operations}. 
While trivial offline, these are costly on embedded UWB tags. For instance, computing the exact amplitude for \cirp takes $\approx\,$\SI{7.9}{\ms} on the \dwmodule.

\boldpar{Amplitude approximation.} 
To minimize execution time, we use the \textit{alpha max plus~beta min algorithm}~\cite{filipAlphaMaxBetaMin}, which approximates $\sqrt{(r^2 + i^2)}$ as \mbox{$\alpha\max(|r|, |i|) + \beta\min(|r|, |i|)$}, with \mbox{$\alpha/\beta = \sqrt{2}+1$}. 
To enable fast, integer-only, on-device computation, we replace the optimal real-valued coefficients with $\alpha{=}5$ and $\beta{=}2$, which closely approximate their ratio.

\begin{figure}[!t]
    \centering
    \vspace{-2.50mm}
    \includegraphics[width=0.53\columnwidth]{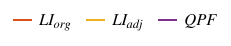}
    \\
    \vspace{-2mm}
    \begin{subfigure}[t]{0.44\columnwidth}
        \centering
        \includegraphics[width=\linewidth]{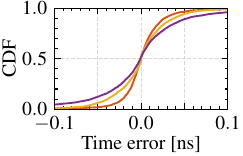}
        \vspace{-5.00mm}
        \caption{CDF}
        \label{fig:cfg-cir-amp_comparison_fp_algos_cdf}
    \end{subfigure}
    \hfill
    \begin{subfigure}[t]{0.434\columnwidth}
        \centering
        \includegraphics[width=\linewidth]{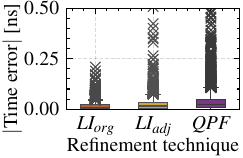}
        \vspace{-5.00mm}
        \caption{Boxplot}
        \label{fig:cfg-cir-amp_comparison_fp_algos_box}
    \end{subfigure}
    \vspace{-1.25mm}
    \caption{Error (in~\SI{}{\nano\second}) induced in the final RX timestamps when using approximated instead of exact amplitude calculation with the FP refinement techniques designed in~\cref{subsec:fprefine}.}
    \label{fig:cfg-cir-amp_comparison_fp_algos}
    \vspace{-3.00mm}
\end{figure}

\boldpar{Evaluation.} We reuse the preamble CIRs collected from
\envitihallway (\cref{subsec:txid}) and run the \emph{entire} CIR
analysis with the exact and approximated techniques.
\cref{fig:cfg-cir-amp_comparison_fp_algos} shows the approximation
error in the final RX~timestamps from our FP~refinement techniques, as
the peak amplitude profile bears the strongest impact on this last
analysis step. The chart confirms that \emph{the approximation error
  is very small across techniques}.  In~\qpf, the most affected
one,~$>$\,92\% 
of RX~timestamps have errors $<\,$\SI{0.1}{\nano\s}, corresponding~to
errors below \SI{1.5}{\centi\meter} for ranging and
\SI{3}{\centi\meter} for \tdoa localization. Still, approximation
brings a \emph{substantial improvement in runtime}, completing in only 
\SI{660}{\micro\s}, \ie \emph{$12\times$ faster than exact
  \mbox{amplitudes}}. Arguably, cm-level inaccuracies are a small
price to pay for the order-of-magnitude overhead reduction enabling
on-device operation.

\newboolean{mergedDMAWin}
\setbool{mergedDMAWin}{true}

\subsection{Limiting and parallelizing CIR extraction}
\label{subsec:bounded}

\noindent
Despite this reduction, the total on-device processing time
(\cref{fig:ctx-costs}) to obtain the final RX timestamps remains high,
$\tproc\seq{\approx}$$\SI{2.5}{\ms}$. Part of this time is
  occupied by the CIR analysis, \ie amplitude calculation, noise level
  estimation, and FP detection and refinement for each
  \lrespi. However, thanks to our techniques, analysis amounts only to
  $\seq{\approx}37$\% of \tproc; the rest of the
  time,\,$\seq{\approx}$$\SI{1.6}{\ms}$, is spent extracting the CIR
  via SPI.

This time is dictated by hardware
and hence
essentially incompressible---\emph{if the entire CIR must be
  acquired}. However, this is necessary only in state-of-the-art
systems, where responder identification hinges on searching the CIR
for the peak-free window \wnoise~(\cref{subsec:txid})---but not when
using our techniques.

\boldpar{From CIR \segs to \sects.} 
First, thanks to our early preamble technique, identifying the peak of
the main responder \lmainr does not require reading \emph{any CIR
  sample} (\cref{subsec:txid}).  The remaining steps of the CIR
analysis determine the FP for all responders \lrespi by operating in
the corresponding CIR segment \ciri. Thus far, we implicitly assumed
contiguous segments covering the entire CIR, as in
existing works (\cref{fig:stscir}). However, this need not be the
case; for each \lrespi, the possible peak positions are concentrated
in a much smaller sample interval, a CIR \emph{sub-segment}, whose
center is easily determined from the expected position of \lmainr and
the known delay \tshift added to transmission by each \lrespi.
Combined, these observations mean that, unlike state-of-the-art
approaches, we can bound CIR extraction to \emph{one small \sect per
  responder}, thus reading only small, non-contiguous CIR portions.

\boldpar{Determining the sub-segment size.} The question becomes how
to determine the size \subciri, in samples, of the \sect to be
extracted for each \lrespi.  The answer depends on how much the peak
of \lrespi can move \wrt the center of the
\sect{}. 
In practice, this hinges on two factors:
    \begin{inparaenum}
    \item how precisely the radio enforces the position of \lmainr in
        the CIR, and
    \item how much the peak of any other responder can move \wrt \lmainr. 
    \end{inparaenum}
    Concerning the former, we empirically verified that our early
    preamble technique results in a variability in the peak position
    of \lmainr of at most $\pm15$ samples.  Therefore, we limited the
    size of the CIR \sect for \lmainr to $\subcir_0\seq{=}30$~samples.
    Concerning the second factor, the peak position of any other
    responder $\lresp_\inonzero$ \wrt \lmainr depends also on the
    propagation time between responder and tag. Thus, \subcirnonzero
    should be set by considering the maximum possible distance \Dmax
    between tag and~$\lresp_\inonzero$.

    When the position of all responders is known, \eg in a fixed
    anchor infrastructure, \Dmax can be easily determined \emph{based
      on purely geometrical considerations} as in
    \cref{subsec:fpdetect}, along with the maximum change in
      peak reception time, $2\Dmax/c$. By dividing the latter value by
      the sample duration (\SI{1.0016}{\ns}), we obtain the maximum
      sample difference \Smax and set
      $\subcirnonzero\seq{=}\subcir_0 + \Smax$ to ensure that the
    peak falls inside the \sect.
  Alternatively, when geometric information is not available but CTX
  exchanges are not infrequent,
  the tag can track peak indexes over
  time 
  and center the CIR \sect{s} to be extracted around the expected peak
  position.

\begin{figure}[!t]
    \centering
    \vspace{-2.75mm}
    \includegraphics[width=0.83\columnwidth]{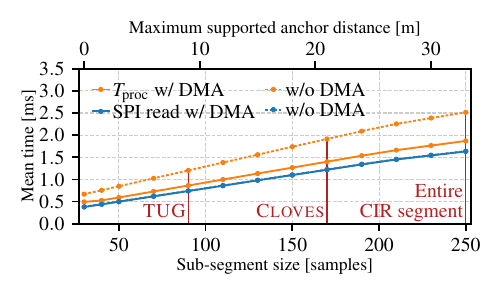}
    \vspace{-3.50mm}
    \caption{The size \subcir of the CIR \sect impacts CIR processing time and maximum distance between anchors \Dmax.}
    \label{fig:runtime_dma_bounded}
    \vspace{-3.75mm}
\end{figure}

\boldpar{Parallelizing operations with DMA.}  Limiting CIR extraction
directly reduces SPI transfer time, but not analysis time which we
already optimized as much as possible. Still, further gains can be
seized by running the analysis \emph{in parallel} with extraction,
exploiting the common DMA functionality offered also by the nRF52833
we use.
DMA allows the SPI read command to fill a specified buffer \emph{without MCU intervention}, freeing the MCU to perform other tasks---from amplitude approximation to FP refinement---\emph{while the CIR is being extracted}.

\evalpar We conduct an experimental campaign to quantify the
benefits of these techniques.  For both testbeds, we consider
  the center tag with the four outermost nodes as responders
(\cref{fig:envs}), and vary \subciri from $\subcir_0\seq{=}30$ to
250~samples, \ie the entire CIR.  As \cirp and \cirs exhibit the same
trends, we focus only on the latter for brevity.

\cref{fig:runtime_dma_bounded} shows 
\begin{inparaenum}
\item the time to only extract the CIR over SPI (blue), and
\item the total time \tproc to extract \emph{and} analyze it (orange),
    \end{inparaenum}
    both averaged over the same 100~repetitions. 
    Results clearly show that using DMA and limiting CIR extraction to
    \sects unlocks significant gains.  In \cloves, using DMA alone
    already yields a 26\% reduction in CIR processing, even with
    $\subcir\seq{=}250$, \ie acquiring the entire CIR; the remaining
    time is dominated by the (incompressible) CIR extraction.
    Using a smaller $\subcir\seq{=}170$, \ie the optimal value based
    on \Dmax, yields an extra 19\% improvement, resulting in an
    overall 45\% reduction \wrt the case without
    DMA. 
    The improvement is even higher in \tug, where the smaller area
    allows a smaller $\subcir\seq{=}90$, yielding a 54\% improvement
    \wrt reading the entire CIR
    with DMA and 66\% without it.


\vspace{-0.75mm}
\section{Putting It All Together: \cloc}
\label{sec:putting_things_together}
\vspace{-0.25mm}

\noindent In~\cref{sec:cloc-design}--\ref{sec:ondevice}, we presented
our techniques to tackle the~main challenges
(\challengeRef{OnDevice}--\challengeRef{SignalSelect}) hindering the
practical adoption of CTX-based systems, and evaluated their
performance and tradeoffs \emph{individually}. In this section, we
build on these individual results and demonstrate the effectiveness of
their \emph{combined} use by presenting and evaluating \cloc: the
first full-fledged CTX-based system enabling \emph{on-device} \dltdoa
localization.

\vspace{-1.25mm}
\subsection{\cloc implementation}
\label{subsec:cloc_internals}

\newcommand{\evalrespfail}{\emph{\resppkt}\xspace}
\newcommand{\evalfpfail}{\emph{FP detection failure}\xspace}
\newcommand{\evaltrfail}{\emph{Triangle inequality check}\xspace}
\newcommand{\evalboxfail}{\emph{Out-of-bounds check}\xspace}

\newcommand{\evalfpfaillp}{failed FP detections\xspace}
\newcommand{\evaltrfaillpshort}{triangle inequality\xspace}
\newcommand{\evaltrfaillp}{\evaltrfaillpshort~checks\xspace}
\newcommand{\evalboxfaillp}{out-of-bounds checks\xspace}

\noindent
\cloc~is~developed~atop~Zephyr~\cite{zephyr}~for~the~\dwplat platform and embeds all techniques developed in~\cref{sec:cloc-design}--\ref{sec:ondevice}. 

\boldpar{Variants under evaluation.}  We showcase different aspects of
our contributions via three \emph{variants} of \cloc
(\cref{tab:techn_chosen}).  \clocp embodies all our core techniques
in~\cref{sec:cloc-design} and~\cref{sec:ondevice} by relying solely on
\cirp, catering for legacy UWB
radios. 
\clocs applies the same techniques but leverages \cirs and our key
switching technique~(\cref{sec:concurrentsts}) to reject unwanted
signals, targeting newer IEEE~802.15.4z
radios. 
Finally, \clocrs is a simple proof-of-concept atop \clocs to
study the performance of the dynamic responder selection enabled by
key switching.
We configure all variants with the best parameters identified
in~\cref{sec:cloc-design}--\ref{sec:ondevice}, and set
$\tshift=\SI{250}{\ns}$ as TX delay.

\begin{table}[!t]
    \centering
    \scriptsize
    \caption{Techniques (\cref{sec:cloc-design}--\ref{sec:ondevice})
      used in the \cloc variants under evaluation
      in~\cref{sec:putting_things_together}.}
    \renewcommand{\arraystretch}{0.85}
    \vspace{-1.00mm}
    \label{tab:techn_chosen}
    \begin{tabular}{@{} r @{\hspace{2.00mm}} l @{\hspace{2.5mm}} c @{}}
    \toprule
    \textbf{Task / Goal}           & \textbf{Technique} & \textbf{Variant} \\
    \midrule
    High-resolution scheduling         & Bidirectional detuning (\cref{subsec:highsched})  & All   \\ 
    Drift estimation                   & Double synchronization (\cref{subsec:highsched}) & All    \\ 
    \lmainr identification       & Early preamble (\cref{subsec:txid})  & All                \\ 
    FP detection                       & Two-stage detector (\cref{subsec:fpdetect}) & All \\ 
    FP refinement                      & \liadj (\cref{subsec:fprefine})      & All                \\
    CIR amplitude calculation & Amplitude approx. (\cref{subsec:fastcir})   & All   \\
    Parallel execution         & DMA (\cref{subsec:bounded})                         & All     \\
    CIR extraction                     & Sub-segments (\cref{subsec:bounded})        & All         \\
    Signal selectivity 
    & STS \& key switching (\cref{sec:concurrentsts}) & \clocs, \clocrs      \\ 
    Dynamic responder selection  & Proof-of-concept (\cref{subsec:cloc_internals}) & \clocrs      \\ 
    \bottomrule
    \renewcommand{\arraystretch}{1.0}
    \end{tabular}    
    \vspace{-3.25mm}
\end{table}


\boldpar{Localization engine.} 
To determine the tag position, \cloc uses known localization techniques that, albeit not being part of our main contribution, are reported next for completeness. 

In \dltdoa systems, the position \mbox{$\posexp_i = [x_i, y_i]^T$} of any anchor (\ie responder \lrespi) is known. In an ideal scenario with
perfect measurements, the \tdoa at the tag position $\posexp$ for a
pair of anchors $(\lresp_i,\lresp_j)$ is
\mbox{$\tdoaexp{ij} = \frac{1}{c}(||\posexp_i-\posexp|| -
  ||\posexp_j-\posexp||)$}, \ie a hyperbola whose foci are the anchor
positions $\posexp_i$, $\posexp_j$ and the tag position \posexp. 
Therefore, to determine the tag position \posexp, it is enough to compute the intersection 
of three~hyperbola.
In practice, however, the \tdoa value measured at the
tag as the difference of the RX~timestamps (\mbox{$\tdoaest{ij} = \trx{i} - \trx{j}$}) is unavoidably affected by an error. 
Hence, instead of identifying the intersection of three hyperbola, the \emph{estimated} tag position \posest should be computed by minimizing the cost function: 
$\posest\seq{=}\text{argmin}_{\posexp}(\tdoaestv - \tdoaexpv)^T(\tdoaestv - \tdoaexpv)$, 
where \tdoaestv and \tdoaexpv are vectors containing
the measured \tdoa values and the ideal, expected ones
at the candidate position \posexp, respectively. 

In \cloc, to solve this cost minimization problem we use a
Gauss-Newton iterative solver~\cite{nocedal2006numerical}, which
refines the position estimate in 5~iterations with a regularization
factor of $0.001$ 
to ensure numerical stability and convergence.  Although three \tdoa
values are sufficient for localization, \cloc uses all six \tdoa pairs
from four anchors to yield more robust 
estimates.

\boldpar{Rejection criteria.} 
To avoid returning invalid position estimates,
\cloc executes a sequential validation pipeline rejecting a
localization attempt at any of the following
stages: 
\begin{itemize}
\item \evalfpfail: no sample with amplitude above threshold \peakthi can be
  detected in a given CIR segment \ciri.
\item \evaltrfail: the RX timestamp of all responders can be computed,
  but their \tdoa values fail the triangle inequality
  check~\cite{xiong15tonetrack}, \ie they yield spatial differences
  exceeding the distance between anchors;
\item \evalboxfail: the final, estimated tag position falls outside the convex hull defined by the anchor positions~\cite{vecchia2019talla}, with 
a spatial tolerance of \SI{0.5}{\m}.
\end{itemize}

\vspace{-0.50mm}
\subsection{Localization performance}
\label{subsec:acc}

\begin{figure}[!t]
    \newlength{\clocMapHeight}
    \setlength{\clocMapHeight}{8cm}
    \centering
    \vspace{-1.00mm}
    \includegraphics[width=0.6\columnwidth]{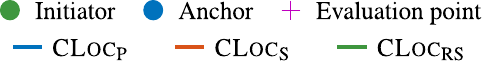}\\
    \vspace{1mm}
    \begin{subfigure}[t]{\columnwidth}
\centering
                \includegraphics[width=\clocMapHeight]{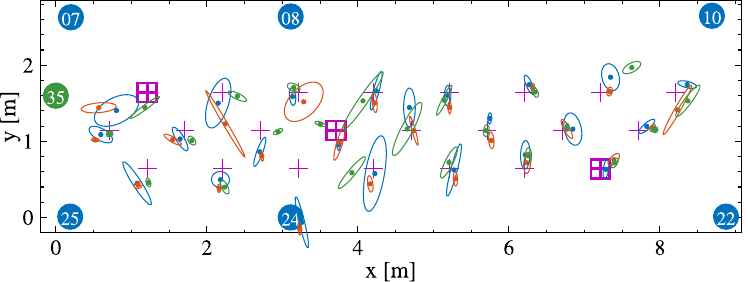} %
        \vspace{-1.00mm}
        \caption{\envitihallway}
    \end{subfigure}\\
\vspace*{2mm}
    \begin{subfigure}[t]{\columnwidth}
\centering
        \includegraphics[width=\clocMapHeight]{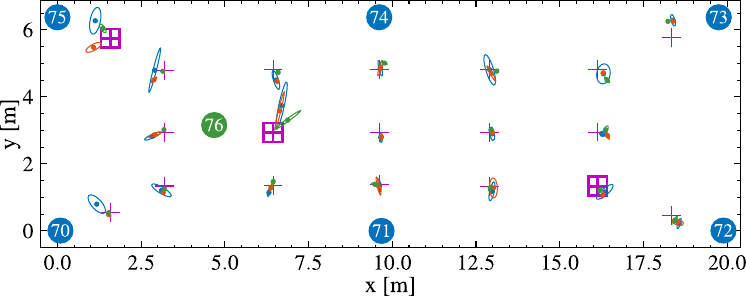}
       \vspace{-1.00mm}
        \caption{\envtuthallway}
    \end{subfigure}
    \vspace{-1.5mm}
    \caption{\cloc-based
      localization. 
      Violet crosses are evaluated tag positions; dots are mean
      position estimates; ellipses are the uncertainty region (\ie 1
      standard deviation) of each estimate.  }
\label{fig:in-action_localization_map}
\vspace{-2.50mm}
\end{figure}

\noindent 
To analyze the reliability and accuracy of \cloc localization, we
perform dedicated experiments in both \tug and \cloves. We begin by
focusing only on the core variants, \clocp and \clocs, then study via
\clocrs how the flexibility of dynamic responder selection may unlock
further improvements.

\boldpar{Experimental setup and methodology.}  In each setup, we use
the four outermost nodes as anchors~(\cref{fig:envs}).  We
explore~23~tag positions in~\tug and 19~in~\cloves (violet crosses
in~\cref{fig:in-action_localization_map}), 
performing 75~localization attempts in each. 
We repeat the same procedure~for 
\clocp and \clocs, and compare their performance in terms of
reliability (\ie number of failed~localization attempts) and accuracy
(\ie absolute position error).

We also study performance under UWB interference, when unintended
transmitters operate on the same channel, and repeat the experiments
with nodes $8,\,24$ in \envitihallway and $71,\,74$ in \envtuthallway
generating UWB traffic interfering with the \resp packets of intended
responders, similar to~\cref{subsec:pre-vs-sts}.

\newcommand{\wint}{(interf.)}
\begin{table}[!t]
  \caption{Rejection rates and ranging error
    for~\cref{fig:cloc_accuracy_rejection}--\ref{fig:clocrs_accuracy_rejection}.}
    \vspace{-.5mm}
    \centering
    \renewcommand{\arraystretch}{.8}
    \setlength{\tabcolsep}{4pt}
    \scalebox{1}{
\begin{tabular}{l|l|c|cccc}
&  \multirow{2}{*}{\textbf{Variant}} & \multirow{2}{*}[1pt]{\shortstack{\textbf{Rejection}\\[-2pt]\textbf{rate [\%]}}} & \textbf{\pc} & \textbf{\psf} & \textbf{\pnt}  & \textbf{\pnn} \\
 & & & \textbf{[cm]}      & \textbf{[cm]}      & \textbf{[cm]}      & \textbf{[cm]}       \\
\toprule
\multirow{6}{*}{\rotatebox{90}{\envitihallway}}
  & \clocp     & 4.1   & 16.6 & 24.8 & 38.6 & 139.6 \\
  & \clocs          & 3.2   & 20.0 & 27.8 & 41.8 & 107.0 \\
 & \clocp \wint   & 52.3  & 22.3 & 40.5 & 71.7 & 153.7 \\
  & \clocs \wint   & 5.2   & 18.1 & 29.9 & 44.9 & 118.1 \\
  \addlinespace[2pt]
  \cline{2-7}
 \addlinespace[2pt]
  & \clocrs       & 2.7   & 17.2 & 24.5 & 32.8 & 103.1 \\  
\midrule
\multirow{6}{*}{\rotatebox{90}{\envtuthallway}}
  & \clocp     & 16.8  & 22.0 & 37.5 & 66.3 & 152.7 \\
   & \clocs          & 3.5   & 22.3 & 37.7 & 49.9 & 89.6  \\
 & \clocp \wint   & 56.9  & 31.8 & 71.5 & 140.9 & 249.8 \\
 & \clocs \wint   & 6.3   & 24.5 & 40.9 & 67.6 & 126.6 \\
  \addlinespace[2pt]
  \cline{2-7}
 \addlinespace[2pt]
 & \clocrs       & 1.9   & 17.1 & 28.6 & 46.3 & 60.6  \\
  \bottomrule
\end{tabular}
    }
    \label{tab:tmpPercentiles}
    \renewcommand{\arraystretch}{1.00}
    \vspace{-1.5mm}
\end{table}

\begin{figure}[!b]
    \vspace{-2.00mm}
    \includegraphics[width=\columnwidth]{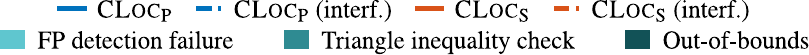}
    \vspace{-3.50mm} \\
    \begin{subfigure}[t]{0.235\textwidth}
        \centering
        \includegraphics[width=0.98817\textwidth]{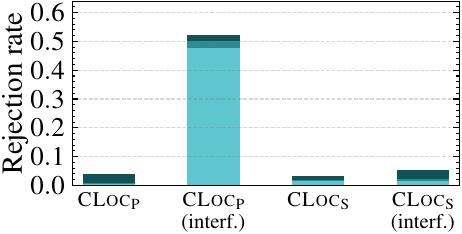}
        \vspace{-5.75mm}
        \caption{Rejection rate (\envitihallway)}
        \label{subfig:rrtug}
                \vspace{1.5mm}
    \end{subfigure}
    \hfill
    \begin{subfigure}[t]{0.235\textwidth}
        \centering
        \includegraphics[width=0.98817\textwidth]{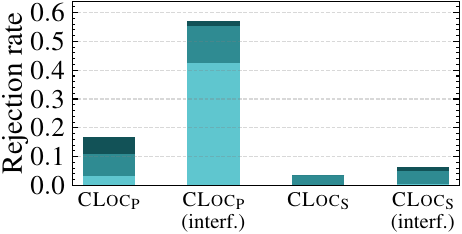}%
        \vspace{-1.75mm}
        \caption{Rejection rate (\envtuthallway)}
        \label{subfig:rrcloves}
        \vspace{1.5mm}
      \end{subfigure}
      \begin{subfigure}[t]{0.235\textwidth}
        \centering
        \includegraphics[width=\textwidth]{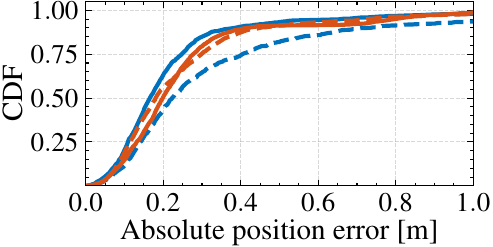}
        \vspace{-5.25mm}
        \caption{Accuracy (\envitihallway)}
        \vspace{+1.25mm}
        \label{subfig:letug}
    \end{subfigure}
    \hfill
    \begin{subfigure}[t]{0.235\textwidth}
        \centering
        \includegraphics[width=\textwidth]{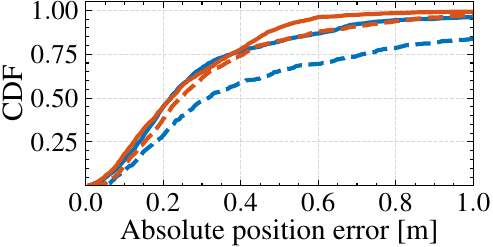}
        \vspace{-5.25mm}
        \caption{Accuracy (\envtuthallway)}
        \vspace{+1.25mm}
        \label{subfig:lecloves}
    \end{subfigure}
    \vspace{-1.75mm}
    \caption{Reliability and accuracy of \clocp and \clocs in \tug and
      \cloves in the presence and absence of interference.}
\label{fig:cloc_accuracy_rejection}
\vspace{-3.00mm}
\end{figure}

\boldpar{Reliability.} 
\cref{tab:tmpPercentiles}~and~\cref{subfig:rrtug}--\subref{subfig:rrcloves}
show that \cloc provides~reliable localization in absence of UWB
interference: the rejection rate remains ${<}4.1$\% in all cases
except for \clocp in \envtuthallway, where it reaches $16.8$\%. In
this case, the culprit is mainly multipath: owing to the large
environment and the greater inter-anchor distances, an incorrect CIR
peak is frequently identified as FP, resulting in a large number of
failed \evaltrfaillpshort and \evalboxfaillp.  However, these
disturbances are remarkably filtered out in \clocs, thanks to the
cleaner \cirs signal, yielding a $4.8\times$ lower rejection rate.

\begin{figure}[!b]
    \vspace{-3.25mm}
    \includegraphics[width=\columnwidth]{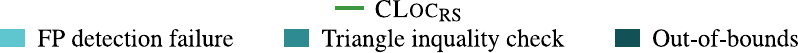}
    \vspace{-3.50mm} \\
    \begin{subfigure}[t]{0.235\textwidth}
        \centering
        \includegraphics[width=\textwidth]{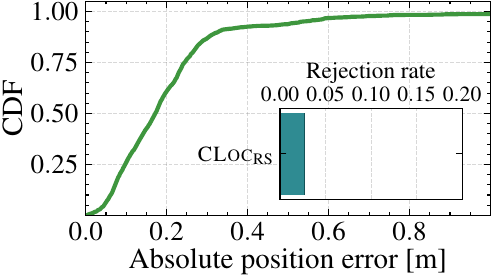}
        \vspace{-5.25mm}
        \caption{\envitihallway}
        \vspace{+1.25mm}
        \label{subfig:rstug}
    \end{subfigure}
    \hfill
    \begin{subfigure}[t]{0.235\textwidth}
        \centering
        \includegraphics[width=\textwidth]{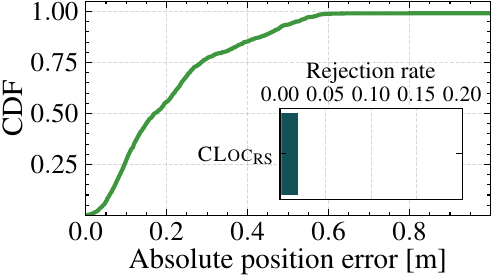}
        \vspace{-5.25mm}
        \caption{\envtuthallway}
        \vspace{+1.25mm}
        \label{subfig:rscloves}
    \end{subfigure}
    \vspace{-2.50mm}
    \caption{Reliability and accuracy of \clocrs.}
\label{fig:clocrs_accuracy_rejection}
\vspace{-3.25mm}
\end{figure}

The difference between the two variants is exacerbated by UWB
interference. With \clocp, more than half of localization attempts are
rejected~in~both environments
(\cref{subfig:rrtug}--\subref{subfig:rrcloves}); interference severely
degrades \cirp, leading to a surge of \evalfpfaillp and \evaltrfaillp.
In contrast, \emph{\clocs is only marginally affected}: rejection
rates increase only by $2.0\%$ (\envitihallway) and $2.8\%$
(\envtuthallway) \wrt the uninterfered case, confirming the markedly
higher robustness of the STS.

Finally, a possible concern is whether our rejection criteria
(\cref{subsec:cloc_internals}) are overly conservative, discarding
many accurate estimates. This is not the case: \emph{none} of the
rejected estimates has error \SI{{<}0.5}{\m}, except for \clocp in
\envtuthallway ($0.5$\%).

\boldpar{Accuracy.} 
Having ascertained the reliability of \cloc, we now inspect its localization accuracy on valid, non-rejected position estimates. 
Figs.~\ref{fig:in-action_localization_map},~\ref{subfig:letug}--\subref{subfig:lecloves}
provide a bird's-eye view of the results, complemented
in~\cref{tab:tmpPercentiles} by commonly used percentile
values; 
we denote the $99^\mathit{th}$ percentile by \pnn and extend this notation to the other percentiles.

In the absence of interference, both variants localize accurately.
In \tug, the median error is \SI{16.6}{cm} for \clocp and \SI{20}{cm}
with \clocs; in \cloves, both yield a median error
${\approx}\,\SI{22}{cm}$.  The same holds at \psf, where both variants
stay well below $1$~m in both environments.
The benefit of STS becomes more evident at higher percentiles,
significantly shortening the error tail.  In \cloves, this is true
already at \pnt, where \clocp incurs an additional error of
\SI{+16.4}{cm} ($+33\%$).  In \tug, the difference remains small
(\SI{<3.2}{cm}) until \pnn, where \clocp yields an additional error of
\SI{32.6}{cm} ($+30\%$).  At the same percentile, the gap widens in
\cloves: \clocp incurs an extra \SI{63.1}{cm} ($+70\%$),
while \clocs stays sub-meter.

Again, interference amplifies the differences between variants. \clocp
shows a marked drop in accuracy: the median error increases by
\SI{+5.7}{cm} ($+34\%$) in \tug and \SI{+9.8}{cm} ($+44\%$) in
\cloves.  Instead, with \clocs the median error is \emph{only
  marginally affected}, remaining within $\seq{\approx}2$~cm (${<}10$\%)
of the uninterfered case in both environments.
Moreover, the gap further increases as percentiles increase. The \pnt
error of \clocp increases \wrt the uninterfered case by \SI{33.1}{\cm}
($+86$\%) in \tug and \SI{74.6}{\cm} ($+113$\%) in \cloves; instead,
with \clocs the \pnt error is only \SI{3.1}{\cm} ($+7.4$\%) and
\SI{17.7}{\cm} ($+35$\%) higher, respectively.  These results confirm
that, under UWB interference, \emph{the use of STS is a necessity} for
reliable and accurate CTX-based \dltdoa localization.

\boldpar{Feasibility of responder selection.} Our STS techniques
enable responder selection: a tag in range of several responders
selects a subset via their keys, using only their signals to
self-position; signals from other responders are filtered out in the
tag~\cirs~(\cref{fig:cfg-keyswitching_message_scheme}), but remain
available to nearby tags. Hence, tags can independently change their
responder sets \emph{at runtime} and the infrastructure can serve
multiple tags at once---improving scalability.
The selection \emph{logic}, \eg choosing anchor pairs closest to the
tag trajectory, or prioritizing responders with good link quality and
line of sight, is application-dependent and outside the scope of this
work. Instead, with \clocrs we evaluate the underlying selection
\emph{mechanism}---dynamic switching of responder sets---confirming
its feasibility and robustness regardless of the logic.

We evaluate \clocrs by setting the two center nodes ($71,\,74$ in
\envtuthallway and $8,\,24$ in \envitihallway) to operate as anchors,
alongside the four outermost nodes.  We then configure the tag to
self-position using only the four~\emph{closest} anchors out of the
six in range, determining at each evaluation point which subset of
anchors to rely on via their ground-truth positions.

\italicpar{Reliability.}  \clocrs is even more reliable than \clocs,
with rejection rates of $2.7$\% (\envitihallway) and $1.9$\%
(\envtuthallway).  By selecting the closest anchors \emph{and
  effectively filtering out the others,} \clocrs benefits from
stronger received signals, which reduce FP detection failures and
improve \tdoa stability, lowering the number of failed
\evaltrfaillpshort and \evalboxfaillp. As for the validation of
rejection criteria, the fraction of rejected estimates ${<}0.5$~m
is \perval{2.3} (\tug) and \perval{0} (\cloves).

\italicpar{Accuracy.}  Results
in~Figs.~\ref{fig:in-action_localization_map},~\ref{fig:clocrs_accuracy_rejection} and~\cref{tab:tmpPercentiles},
show that \clocrs clearly outperforms \clocs (already shown to
outperform \clocp), thanks to two complementary effects.  First,
selecting the closest anchors yields stronger received signals,
improving FP detection and \tdoa stability.  Second, higher anchor
density often translates to a more favorable constellation geometry,
reducing errors caused by, \eg geometric dilution of precision.
Improvements are particularly marked in \cloves, where the median
error drops by $5.2$~cm (\perval{-23}) and the p99 by $29$~cm
(\perval{-32}), \emph{despite simultaneous transmissions of the other
  two anchors acting as interferers,} confirming that \clocrs
successfully and reliably suppresses their impact.

\newcommand{\pointTwoD}[2]{$(\SI{#1}{\m}, \SI{#2}{\m})$}

\vspace{-1.50mm}
\subsection{On-device runtime and energy consumption}
\label{subsec:on-device_localization}
\vspace{-0.25mm}

\noindent 
Besides being reliable and accurate, \emph{on-device} \cloc operation
is fast and energy-efficient. We directly \emph{quantify}, for the
first time, these CTX benefits against a conventional system.

\boldpar{Experimental setup and methodology.} 
We compare \cloc against conventional
staggered~\dltdoa~\cite{dltdoa,crazyflie} (\cref{fig:dl_tdoa}),
hereafter \emph{\nonCtxDLTDoA} for simplicity.
\cref{fig:ctx-costs} illustrates the key difference between the two
over the period \truntime from reception of the first (and, in \cloc,
\emph{only}) \resp packet until acquisition of the four RX timestamps
needed for \tdoa localization.  In \nonCtxDLTDoA, the tag is mainly
occupied receiving multiple \resp packets and handling their payloads;
\cloc eliminates this need, devoting the time after its single
reception to CIR extraction and analysis, whose duration is denoted
\tproc.

We compare the performance of \clocp and \clocs against a reference
\nonCtxDLTDoA system using variants with preamble or STS. A key
parameter of the latter is the \emph{minimum} staggering interval
between two consecutive \resp packets guaranteeing reliable
operation. We measure the interval at the \rmarker of the \resp
packet, determining its minimum value as \SI{307}{\us} when using the
preamble and \SI{423}{\us} with the STS. In the latter case, staggered
employs a 64-symbol STS, ensuring a fair comparison with \cloc, which
processes 128~STS~symbols across two responder pairs.
For both systems, we estimate the time spent
by the radio in each state from the known fixed packet size and a
\SI{10}{\us} RX~guard time, then derive energy consumption based on
these timings, the processing ones measured as
in~\cref{subsec:bounded}, and the nominal instant currents of the
\dwradio and nRF52833
MCU~\cite{qorvoDW31110datasheet,nrf52833current}.

\begin{table}[!t] 
\newcolumntype{Y}{>{\centering\arraybackslash}X}
\caption{On-device runtime and energy consumption of conventional \nonCtxDLTDoA
  \dltdoa \vs \cloc.}
\vspace{-1.25mm}
\label{tab:ondev}
\centering
\renewcommand{\arraystretch}{0.88}
\footnotesize 
\resizebox{\columnwidth}{!}{
    \newlength{\colSep}
    \setlength{\colSep}{3.5mm}
\begin{tabular}{@{\hspace{1.5mm}} c @{\hspace{1mm}} c @{\hspace{\colSep}} c @{\hspace{\colSep}} c @{\hspace{\colSep}} c @{\hspace{\colSep}} c @{\hspace{\colSep}} c @{\hspace{1.5mm}} }
    \toprule
    & \multicolumn{2}{c}{\textbf{\nonCtxDLTDoA}} & \multicolumn{2}{c}{\textbf{\cloc (\envitihallway)}} & \multicolumn{2}{c}{\textbf{\cloc (\envtuthallway)}}\\
    \cmidrule(lr){2-3}
    \cmidrule(lr){4-5}
    \cmidrule(lr){6-7}
                                    & \textbf{Preamble} & \textbf{STS} & \textbf{\clocp} & \textbf{\clocs} & \textbf{\clocp} & \textbf{\clocs} \\
    \midrule
    \tproc [\SI{}{\ms}]    & -                 & -            & 0.809 & 0.792 & 1.226           & 1.200  \\    
    \truntime [\SI{}{\ms}] & 1.063             & 1.476        & 0.909 & 1.023 & 1.326           & 1.431  \\
    Energy [\SI{}{\uJ}]    & 134.1             & 203.7        & 80.4  & 109.9 & 109.9           & 138.8  \\
    \bottomrule
    \renewcommand{\arraystretch}{1.00}
\end{tabular}
}
\vspace{-3.50mm}
\end{table}

Finally, in staggered, runtime and energy consumption are unaffected
by tag position---unlike in \cloc. Therefore, we explore 3~tag
positions in each testbed
(\includegraphics[width=2.6mm]{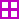} in
\cref{fig:in-action_localization_map}), representative of different
anchor-tag geometries.

\boldpar{Runtime and energy consumption.}  \cref{tab:ondev} shows the
on-device runtimes.  In \cloc, the duration \tproc of CIR processing
depends on the environment, due to our bounded
analysis~(\cref{subsec:bounded}): the area in \tug~is smaller than in
\cloves, yielding a smaller \subcir and shorter \tproc and \truntime
(up to \perval{-34.0} and \perval{-31.4}, respectively) for both \cloc
variants.
As for \nonctx, when using the preamble it is $17$\% slower than
\clocp in \tug, but $20$\% faster in \cloves.  Instead, when using STS,
\nonCtxDLTDoA is \emph{always slower} than \clocs, respectively by
$44\%$ and $3\%$.  The reason is that
\begin{inparaenum}
\item packets are longer, as they must embed the STS, and 
\item this overhead is incurred only once \emph{per position} in \cloc
  but once \emph{per anchor} (\ie four times) in \nonCtxDLTDoA.
\end{inparaenum}
As a consequence, \truntime increases only marginally from preamble to
STS in \cloc ($+13$\% and $+8$\%), whereas it increases
significantly in \nonCtxDLTDoA ($+39$\%).

As for energy consumption, both \cloc variants markedly outperform
their \nonCtxDLTDoA counterparts (\cref{tab:ondev}), since \nonctx's
communication overhead exceeds \cloc's CIR
processing~(\cref{fig:ctx-costs}).  With the preamble, \nonCtxDLTDoA
uses $+67$\% more energy than \clocp in \tug and $+22$\% in
\cloves. The gap widens with STS, where \nonCtxDLTDoA uses $+85$\% and
$+47$\% more energy than \clocs, respectively.

\boldpar{Localization and achievable update rate.}
The RX timestamps output by staggered and \cloc are finally input to
the same localization engine (\cref{subsec:cloc_internals}), which
determines the tag position in $1.82\,\pm\,0.01$\,ms and consumes
$30.6\,\pm\,0.2$\,$\SI{}{\uJ}$.  Adding this fixed contribution to the
values in \cref{tab:ondev} yields the \emph{total} time and energy
consumption for a single position sample, \ie from the reception of
the first \resppkt until the tag position is determined.  For \clocp,
the total time amounts to~\SI{2.75}{\ms} (\envitihallway)
and~\SI{3.14}{\ms} (\envtuthallway), with a corresponding energy
consumption of~\SI{111.2}{\micro\joule}
and~\SI{140.3}{\micro\joule}. For \clocs, the total time is
\SI{2.86}{\ms} (\envitihallway) and \SI{3.25}{\ms} (\envtuthallway)
with an energy consumption of \SI{140.7}{\micro\joule} and
\SI{169.4}{\micro\joule}, respectively.
%
%
These results yield an overall update rate of~\SI{363}{\Hz} (\tug)
and~\SI{318}{\Hz} (\cloves) for \clocp, and \SI{349}{\Hz} and
\SI{307}{\Hz} for \clocs, respectively.  Therefore, in addition to
high energy efficiency, \cloc also enables very high update rates.

\boldpar{Improving further.} This already remarkable performance can
be further increased, \eg in applications involving \emph{tracking},
by letting the tag exploit knowledge from previous samples.  For
example, after the tag establishes its position via the associated
FP~indexes, in the next rounds it can focus only on the CIR portions
around them. This would yield sub-segments even smaller than the
already small ones we currently use, amplifying the advantages in CIR
extraction and processing outlined in~\cref{subsec:bounded}, thus
further reducing runtime and energy consumption at once.
Moreover, when tracking occurs at a relatively high update rate, the
corresponding changes in position are likely small and few iterations
can be afforded in the \tdoa solver (\cref{subsec:cloc_internals})
without compromising accuracy, significantly reducing the impact of
position computation.


\section{Conclusions}
\label{sec:ending}

The idea of UWB concurrent transmissions for ranging and localization
was first proposed and demonstrated in~2018. Since then, several
research works built upon the initial results. Still, none offer
\emph{on-device} operation, let alone quantify its runtime and energy
consumption benefits, confining CTX-based ranging and localization to
\emph{offline} operation, thus significantly hampering practical
applicability and potential impact.

In this paper, we contribute new CTX techniques and redesign existing
ones from the ground up, enabling not only their efficient,
\emph{entirely} on-device execution, but also significantly
improving their performance \wrt existing state-of-the-art
  baselines.
Moreover, by exploiting---for the first time in the
literature---the \ieeez STS within CTX operation we increase
reliability and offer the novel capability of dynamic responder
selection allowing distinct CTX~exchanges to occur
simultaneously, increasing scalability.

Finally, we contribute \cloc, a prototype embodying \emph{all} of the
techniques above. \cloc is not only the first on-device concurrent
\dltdoa system, but also enables a \emph{quantitative} analysis of the
benefits CTX unlocks \wrt a conventional staggered solution---up
to \perval{44} slower and \perval{85} more energy-hungry depending on
the target configuration and environment.

We believe our techniques, and the prototype embodying them, represent
a significant advancement, laying the groundwork for future systems to
fully exploit the potential of CTX-based UWB ranging and localization.


\section*{Acknowledgements}
\label{sec:acks}

\noindent
\looseness=-1
This work was partially supported by:
\begin{inparaenum}
\item the European Union---Next Generation EU under the Italian National
  Recovery and Resilience Plan~(NRRP), Mission~4, Component~2,
  Investment~1.3, CUP~B53C22003970001, partnership on
  ``Telecommunications of the Future'' (PE00000001---program
  ``RESTART'');
\item the project SPACE IT UP! ASI contract n.~2024-5-E.0 CUP Master
  n.~I53D24000060005 funded by the Italian Space Agency (ASI) and
  the Ministry of University and Research (MUR);
\item the BMK, BMAW, and FFG, Contract No. 881844: ``Pro²Future''
  (Products and Production Systems of the Future), within the
  ENHANCE-UWB project; 
\item the SPiDR (``Secure, Performant, Dependable, and Resilient
  Wireless Mesh Networks'') and SPIDR2 (``Secure, Performant,
  Intelligent, Dependable, Reliable, and Resilient Wireless Systems'')
  projects funded by the Technology Innovation Institute.
\end{inparaenum}

\boldpar{AI disclosure statement.} Claude (Anthropic) was used only to
shorten portions of the text and to revise
Figs.~\ref{fig:fpnew_parametrization}
and~\ref{fig:fpdetection_evaluation}, all produced by the authors; no
new content (text, figures, images, or code) was generated with
AI. The authors reviewed all changes and take full responsibility for
this article.


\bibliographystyle{IEEEtran}
\bibliography{refs_s}

\end{document}